\documentclass[12pt]{article}

\usepackage{packages}

\usepackage{simeontex}

\usepackage{mathptmx}
\usepackage{anyfontsize}
\usepackage{t1enc}

\usepackage{definitions}

\newcommand{\marrow}[5]{%
    \fmfcmd{style_def marrow#1
    expr p = drawarrow subpath (1/4, 3/4) of p shifted 6 #2 withpen pencircle scaled 0.4;
    label.#3(btex #4 etex, point 0.5 of p shifted 6 #2);
    enddef;}
    \fmf{marrow#1,tension=0}{#5}}

\newcommand{\Marrow}[6]{%
    \fmfcmd{style_def marrow#1
    expr p = drawarrow subpath (1/4, 3/4) of p shifted #6 #2 withpen pencircle scaled 0.4;
    label.#3(btex #4 etex, point 0.5 of p shifted #6 #2);
    enddef;}
    \fmf{marrow#1,tension=0}{#5}}

\preprint{\texttt{IPMU15-0040}}

\newcommand{\OfficialTitle}{On the CFT Operator Spectrum \\ at \\ \vskip.1in  Large Global Charge}

\title{
  {\Huge\textbf{
      \OfficialTitle}}
}

\hypersetup{pdfauthor={Simeon Hellerman and Domenico Orlando and
    Susanne Reffert  and Masataka Watanabe},pdftitle={\OfficialTitle}}

\author{%
  \begin{minipage}{.8\linewidth}
    \vspace{1cm}
    \begin{center} 
      {\small 
        \textbf{Simeon Hellerman}\textsuperscript{1}, 
        \textbf{Domenico~Orlando}\textsuperscript{2,3},
        \textbf{Susanne~Reffert}\textsuperscript{4} and
        \textbf{Masataka Watanabe}\textsuperscript{1}}
    \end{center}
    \vspace{1cm}
    \authorBlock{1}{Kavli Institute for the Physics and Mathematics of the Universe\\The University of Tokyo.  Kashiwa, Chiba  277-8582, Japan}
    \authorBlock{2}{\textsc{lptens} -- \textsc{umr cnrs} 8549, 24, rue Lhomond, 75231 Paris, France} 
    \authorBlock{3}{\textsc{ipt} Ph. Meyer, 24, rue Lhomond, 75231 Paris, France}
    \authorBlock{4}{\textsc{itp -- aec}, University of Bern, Sidlerstrasse 5, 3012 Bern, Switzerland}
  \end{minipage}
}

\renewcommand{\url}[1]{#1}

\date{}

\begin{document}

\begin{titlepage}

  \maketitle
  \thispagestyle{empty}
  %

     \abstract{\normalfont \noindent
    We calculate the anomalous dimensions of operators with large global charge $J$ in certain strongly coupled conformal field theories in three dimensions, such as the \(O(2)\) model and the supersymmetric fixed point with a single chiral superfield and a $W = \Phi^3$ superpotential.  Working in a $1/J$ expansion, we find that the large-$J$ sector of both examples is controlled by a conformally invariant effective Lagrangian for a Goldstone boson of the global symmetry.  For both these theories, we find that the lowest state with charge $J$ is always a scalar operator whose dimension $\Delta_J$ satisfies the sum rule
   \[ J\sqd  \Delta_J - \left( \tfrac{J^2}{2} + \tfrac{J}{4} + \tfrac{3}{16} \right) \Delta_{J-1}
- \left( \tfrac{J^2}{2} - \tfrac{J}{4} + \tfrac{3}{16} \right)
\Delta_{J+1} =  0.035147  \ , \]
up to corrections that vanish at large $J$.  The spectrum of low-lying excited states is also calculable explicitly:  For example, the second-lowest primary operator has spin two and dimension $\Delta\ll J + \sqrt{3}$.
In the supersymmetric case, the dimensions of all half-integer-spin operators lie above 
the dimensions of the integer-spin operators by a gap of order $J\uu{+\hh}$.  The propagation
speeds of the Goldstone waves and heavy fermions are ${1\over \sqrt{2}}$ and $\pm \hh$
times the speed of light, respectively.  These values, including the
negative one, are necessary for the consistent realization of the superconformal symmetry at large $J$.
      }
  \vfill

\end{titlepage}

\restoregeometry
\setcounter{tocdepth}{2}
\setcounter{secnumdepth}{2}
\tableofcontents
\newpage

\section{Introduction}

\heading{Motivation}

Most \acp{cft} lack nice limits where they become simple and solvable.  Some of them are just in the middle of coupling-constant space, and all anomalous dimensions are of
order $1$, and the \textsc{ope} is just intrinsically complicated, and no parameter of the theory can be dialed to a simplifying limit.

Even in such cases there
may sometimes be sectors of the theory where anomalous dimension and \textsc{ope} coefficients simplify.  Many examples of this type are known, where the simplifying limit involves taking large quantum 
numbers, $J$, under rotational or internal global symmetries. 

The simplifications of such limits at large $J$ have played
a role in the correspondence between strings in holographic spacetime
and single-trace operators in planar gauge
theory~\cite{Berenstein:2002jq}.  The large-$J$ limit has also
simplified the analysis of the high-spin spectrum in
\ac{cft}~\cite{Fitzpatrick:2012yx,Komargodski:2012ek, Alday:2013cwa, Alday:2015eya} and the worldsheet theory in the confining regime~\cite{Hellerman:2014cba}.\footnote{See also~\cite{Baker:2002km,PandoZayas:2003yb} for earlier
attempts at an analysis in the same limit.}

It would be desirable to develop a more general understanding of the simplification
of \ac{cft} at large global symmetry quantum numbers.  First of all, in theories that are strongly
coupled, we should exploit any analytic tools available to gain information about the operator spectrum.
Second of all, such simplifications in the spectrum as have been observed in~\cite{Komargodski:2012ek,Kaviraj:2015cxa,Kaviraj:2015xsa} appear to be fundamental to the structure of \ac{cft} itself as understood through the conformal bootstrap. These simplifications 
have been derived using the abstract rules of the bootstrap rather than any sort of Lagrangian formulation.\footnote{For progress on the simple 3D conformal theories using modern bootstrap methods, see for example~\cite{Rattazzi:2008pe,Rattazzi:2010yc,ElShowk:2012ht,El-Showk:2014dwa,Kos:2015mba}.}  It is therefore quite intriguing that the form of the asymptotic spectrum at large
spin is quite reminiscent of the results we derive in this paper for large global internal symmetry quantum numbers.\footnote{We thank Kallol Sen and Aninda Sinha for discussions and correspondence on this point.}

\heading{Summary of methods}

In this paper we will illustrate the simplification of \ac{cft} at large global symmetry quantum numbers with two simple examples of strongly coupled fixed points in
three dimensions:  The critical point of the $O(2)$ model~\cite{Wilson:1971dc} (or XY model), and the ${\cal N} = 2$ superconformal fixed point of the Wess--Zumino model with a single chiral superfield and a $W = \Phi^3$ superpotential.\footnote{This theory can be shown to flow to a nontrivial fixed point by various means, such as the computation of the two-point function of the $R$-current in the infrared~\cite{Barnes:2005bm, Jafferis:2010un}.} We treat these theories by quantizing them on a 2-sphere of radius \(R\) and calculating their operator dimensions \emph{via} radial quantization.  

In the limit of large $J$, there is a large hierarchy between the radius $R$ of the sphere and the length scale set by
the charge density, $\r\uu{-\hh} \sim J\uu{-\hh} \cc R$.  We can then consider the \ac{cft} as a Wilsonian effective action at a renormalization group fixed-point with cutoff $\L$ and exploit the cutoff-independence of the dynamics to take
\bbb
{1\over R} \muchlessthan \L \muchlessthan \sqrt{\r} = {{\sqrt{J}}\over {2R\sqrt{\pi}}} \ .
\een{TripleHierarchyZero}
which implies $J \muchgreaterthan 1$.
   In the regime $\L \muchlessthan \sqrt{\r}$, it turns out that the effective action is weakly coupled
   and under perturbative control, with an expansion in powers of $\r\uu{-1}$, with numerators
   given by derivatives of $\r$ and powers of the cutoff $\L$.  When we quantize the theory, the leading approximation to
   any quantity is given by the leading large-$\r$ term in the action.  Sub-leading corrections are generated by quantum loops and by explicit vertices with more negative powers of $\r$ in the operators.  
   
   In the Wilsonian action, only non-negative powers of $\L$ appear, and these are small in the
   limit \rr{TripleHierarchyZero}.  Furthermore, the underlying conformal invariance of the theory means that the $\L$-dependence cancels in all physical observables, order by order in the ${1\over J}$ expansion.  The $\L$-dependent terms are scheme-dependent.  They play little role in the dynamics of the large-$J$ theory other than to restore quantum scale invariance in the effective Lagrangian description and to cancel the $\L$-dependence in amplitudes.  We will explain how the $\L$-dependent terms may be calculated algorithmically order by order
   from the $\L$-independent terms for any given form of the cutoff.
   
   \heading{Summary of results}
     
In the three-dimensional examples we consider, we find that leading large-$J$ behavior
of the dimension $\Delta\ll J$ of the lowest operator with global charge $J$ goes as 
\bbb
\D\ll J \propto J\uu{+{3\over 2}}
\een{UniversalLargeJScaling3D}
 at large $J$.  This is true both in the $O(2)$ model and also in the $W= \Phi\uu 3$ model.  Note
that the dimension in the latter case does \emph{not} go as $\Delta=J+O(J^0)$, as might naively have been expected based on supersymmetric considerations. Despite the
presence of a \textsc{bps} bound and multiplet-shortening condition in this ${\cal N} = 2$ theory, the operators
$\phi\uu J$ do \emph{not} saturate it, even approximately: The lowest state in the large-$J$ sector
is parametrically far above the supersymmetric bound.

We can also compute the sub-leading terms in the expansion.  In both the $O(2)$ model and the $W = \Phi\uu 3$ model, the lowest dimension in the large-$J$ sector has the expansion
\begin{equation}
\Delta\ll J =   c\ll{3\over 2} \cc J^{3/2} + c\ll{1\over 2} \cc J^{1/2} - 0.0937256 + O(J\uu{-{1\over 4}})
\end{equation}
at large $J$.
The coefficients $c\ll{3\over 2}$ and $c\ll{1\over 2}$ are related to the coefficients of the leading and first sub-leading terms in the large-$J$ effective Lagrangian and 
we do not at present know how to calculate them from first principles.  They may differ
between the supersymmetric and bosonic models.  The order $J\uu 0$ term, on the other hand,
is calculable and common to the two models, and to any other model described by the same large-$J$ universality class. 

\heading{Dimensional analysis and large-$J$ scaling}

We note here that the \emph{leading} large-$J$ scaling can be deduced immediately on
dimensional grounds without the need for the methods developed in this paper.
At large charge, the charge density $\r$ and energy density ${\cal H}$ are
homogeneous and semiclassical on distance scales between $R$ and $\r\uu{-\hh}$.  In
this range of scales, the two densities must obey a local and scale-invariant relationship,
which in three dimensions can only be of the form
\bbb
{\cal H} \sim \r\uu{+{3\over 2}}
\een{UniversalLargeJScaling3DLocal}
at large $J$, leading to the scaling \rr{UniversalLargeJScaling3D}.

The only exceptions to the rule \rr{UniversalLargeJScaling3DLocal} in three dimensional \ac{cft} are theories that have a vacuum manifold of exactly flat directions, such as a free complex scalar or a supersymmetric theory with a quantum mechanically supersymmetrically protected moduli space.  For such theories, the size of the sphere
is \emph{never} irrelevant, because in the absence of the conformal coupling term of the scalar fields to the Ricci curvature, the spectrum of the Hamiltonian would collapse and become continuous.  Such theories do indeed have continuous spectra on flat spatial slices such as $T\uu 2$, and therefore
the Ricci curvature ${\tt Ric}\ll 3$ must enter into the leading term in any local relationship between
the charge density and energy density. Such theories obey ${\cal H}\sim ({\tt Ric}\ll 3)\uu{+\hh} \cc \r$ for the ground state at large $J$.

\ac{cft} with good spectra on flat slices, on the other hand -- by which we mean, theories with discrete energy levels when quantized on $T\uu 2$ -- are of
a more generic character, and include all known interacting non-supersymmetric theories,
and even many supersymmetric ones.  The critical $O(2)$ model, the $W = \Phi\uu 3$ model, three-dimensional super-\textsc{qed} and the $\IC\IP(N)$ model~\cite{Dyer:2015zha}, non-supersymmetric Chern--Simons theories with matter~\cite{Aharony:2012nh,Aharony:2015pla}, and even theories of free fermions all fall into this category.  The scaling of the operator dimension
with global charge in such theories is \emph{always} $J\uu {3\over 2}$.  This scaling is
directly visible, for instance, in the theory of a free relativistic fermion, where
the $J\uu {+{3\over 2}}$ is just the ground state energy of a Fermi surface on the sphere.
The same scaling also appears in interacting three-dimensional theories, such
as Chern--Simons theory with matter, at large magnetic flux number.\footnote{We thank Ethan Dyer and Mark Mezei for
discussions and correspondence on their results for this example and related work by them~\cite{Dyer:2013fja,Dyer:2015zha,DyerStanfordApril}.}

Despite the automaticity of the leading-order $J\uu{+{3\over 2}}$ scaling itself, the power laws appearing in the sub-leading large-$J$ corrections do not follow directly from dimensional analysis, 
nor does the computability of the coefficients of those corrections
in perturbation theory.  Rather, the structure of the sub-leading corrections follows from a renormalisation-group analysis that may depend on the details of the theory.

\heading{Outline}

The plan of the paper is as follows.    
In Section~\ref{sec:BigJRG}, we discuss the RG flow and large-$J$ perturbativity based on a toy model. In Section~\ref{sec:O2-3D}, we apply the lessons we have learned to the full $O(2)$ model in three dimensions and discuss its large-$J$ analysis and study its reduction to Goldstones after integrating out the $a$ field. We go on to discuss the classification of operators in the conformal Goldstone action. In Section~\ref{sec:phi3}, we move on to the supersymmetric $W=\Phi^3$ model. We perform its RG analysis at large $J$, decouple the fermions and compute the large-$J$ expansion of operator dimensions using radial quantization. In Section~\ref{sec:observables}, we work out the energy of the excited states and see which states are primaries and which are descendants. In Section~\ref{sec:other-models}, we go on to discuss other models in three and four dimensions.
In Section~\ref{sec:conclusions}, we present our conclusions and point out interesting future directions.

 \section{RG flow with a dimensionful vev}
\label{sec:BigJRG}

We would like to
investigate to what extent the operator spectrum
simplifies and becomes calculable at large $J$ in generic \ac{cft},
particularly in models for which there are no other tools or
tunable parameters of the Hamiltonian.  A paradigmatic example of such a theory is the
Wilson--Fisher point of the $O(2)$ model in three dimensions, whose ultraviolet definition is simply
a complex scalar field with potential term $V = \cc {{g\sqd}\over{12}} \cc |\phi|\uu 4$.
\def\naive{naive\cc}

\def\okcol{Blue}
\def\ok{{\color{\okcol} {\checkmark}}~}



%
  

On the one hand one expects the spectrum of this model to be under perturbative control for large $J$: The loop-counting parameter for a process with characteristic scale $R$ and initial and final state $\kket{J}$ is $y = g^2 R/J$. On the other hand, the energy of the $J\uth$ state on a sphere of radius $R$ scales as $J/R$ and becomes so large as to exceed the scale where the ultraviolet physics decouples if we require $y \muchlessthan 1 $, \emph{i.e.} $J/R \muchgreaterthan g^2$. This would suggest that large-$J$ perturbativity is not useful for the computation of operator dimensions in the interacting \ac{cft}.  Such a conclusion is too pessimistic. We shall see how to compute certain quantities in the large-$J$ sector in a
  controlled fashion.  
  
  We first turn to the analysis of the renormalization group equation for this system, in states of large $J$.  We will begin with a toy model of the RG flow, that can be thought of as the $O(2)$ model
  where only the flow of one particular operator is retained.  After understanding the general behavior
  of the toy model, we will return to the full $O(2)$ model and
  analyze its behavior at large charge density.  Then, we will solve the full RG equations at the fixed point, in the large-$J$ expansion.  The solution in this limit reduces to an effective Lagrangian
  that is classically scale invariant, plus small quantum corrections suppressed by positive powers
  of the ratio of the cutoff to the square root of the charge density.
  
This Lagrangian explicitly realizes the conformal symmetry of the underlying \ac{cft}, even
    while strongly spontaneously broken by the charge density itself.  We then quantize this theory
    on the unit $S\uu 2$ spatial slice, and calculate the energy.  This gives the value of the dimensions of operators at large $J$ via radial quantization according to the state-operator correspondence.  The energy of the lowest state at large $J$ can be calculated in an asymptotic expansion in inverse powers of $J$ in terms of our effective Lagrangian.  Certain terms in the expansion depend on
    the coefficients in the large-$J$ Lagrangian, while others are universal and theory-independent altogether.

\subsection{Toy model of the RG flow}

To understand the effect of large $J$ on the renormalisation group flow, let us write a toy
model for the renormalization group equation for the effective action of the $O(2)$ model.
In this model, we discard renormalizations of operators with derivatives and also renormalizations
of the mass term and other terms in the potential itself, other than the quartic coupling.

We define this model in the spirit of the toy model in~\cite{Polchinski:1992ed}: We take the structure of a true exact Wilsonian RG flow for a real scalar field, truncated to the flow
for the quartic coupling.
We discard the cutoff dependence of the mass term, the kinetic term, and all higher-derivative terms.
Though this is an uncontrolled approximation of the full RG flow, as in the toy model of~\cite{Polchinski:1992ed}, it illustrates a key behavior
of renormalisation group flow in a situation where the infrared quantum scale invariance is strongly spontaneously broken by the expectation value of a field.

The ultraviolet Lagrangian will be that of a three-dimensional real scalar with a quartic potential:
\begin{equation}
{\cal L}\ls{\rm micro} = -\hh \cc (\pp  a)\sqd - {{g\sqd}\over{12}}\cc a\uu 4\  .
\end{equation}
\def\uprm#1{^{({\rm #1})}}
\def\eff{\text{eff}}
\def\gseff#1{g\sqd\ls{\eff}({#1})}

\heading{Conformal behavior for $\L \muchlessthan g\sqd$}

We integrate out modes and lower the cutoff, which is recorded by
a $\L$-dependent evolution of the effective Lagrangian ${\cal L}\ls \L$, of which we keep track of only the renormalized quartic coupling $\gseff{\L}$.  At $\L \sim g\sqd$,
the quantum corrections to the
coupling $\gseff{\L}$
are of the same order as $g\sqd$ itself.  At that point, the coupling $\gseff{\L}$ quickly
reaches its attractive\footnote{In the full field theory, the fixed point is not fully
attractive; one must fine-tune the mass term for $a$ in order to stay on the fixed point.  In our
model, however, we have simply truncated out the quantum running of the infrared-unstable coupling $m\sqd a\sqd$ along with all the higher-derivative couplings we don't want to keep track of.}
 fixed-point value in units of $\L$, after which point it obeys
\begin{equation}
 {{\gseff{\L\ll 1}}\over{\L\ll 1}} = {{\gseff{\L\ll 2}}\over{\L\ll 2}}\ , \llsk\L\ll{1,2} \muchlessthan g\sqd\ ,
\end{equation}
 or equivalently
 \begin{equation}
 \gseff{\L} = h\cc \L \ , \llsk\L \muchlessthan g\sqd\ ,
\end{equation}
 where $h$ is a dimensionless coupling whose numerical value is 
 determined by the fixed point equation.
 
 \heading{\textsc{vev} for the $a$ field as an infrared cutoff}

We can calculate the effective Wilsonian action for any value we like of the modes
below the cutoff.  In particular, we would like to give a nonzero value to the
constant mode of the scalar field $a$.  We assume this value is far below the scale set by
the ultraviolet scale, so $a \muchlessthan g$.  This expectation value induces a mass 
\bbb
m\ll a\sqd = g\sqd\ls{\eff}\cc a\sqd
\een{PhiFlucMass}
 for the $a$-fluctuations.  When we integrate down to momentum modes comparable to or below $m\ll a$, 
the evolution of the couplings stops, because modes of $a$ no longer
make large contributions to the running when they are below the gap.  Fluctuations of such modes renormalize the
couplings only with a suppression by positive powers of ${{\L}\over{m\ll a}}$.  Therefore,
at approximately 
\bbb
\L\ll{{\rm mass}} = m\ll a 
\een{IRCutoff}
the evolution of $g\uprm{\eff}$ comes to a halt.
Combining equations \rr{PhiFlucMass} and \rr{IRCutoff},
the infrared value of the coupling satisfies
\begin{gather}
  g\sqd\ls{\eff}(\L\ll{\rm mass}) = h\sqd \cc a\sqd\ , \\
\L\ll{\rm mass} = m\ll a = h \cc a\sqd\ .
\end{gather}
Then the effective potential below $\L\ll{\rm mass}$, which is $V\ls{\eff}(a) = {1\over{12}}   g\sqd\ls{\eff}(\L\ll{\rm mass}) \cc a\uu 4$ becomes 
\begin{equation}
  V\ls{\eff}(a) = {{h\sqd}\over{12}} \cc a\uu 6\ .
\end{equation}
This form for the low-energy effective potential could have been inferred just on the basis of dimensional analysis, but it is instructive to see how the sextic potential emerges from the structure
of the renormalization group.

For $\L < \L\ll{\rm mass} = h a\sqd$, the effective action asymptotes to a constant as
$\L\to 0$.  Despite this,
the theory is still conformally invariant; the conformal invariance is simply broken spontaneously
by the expectation value of $a$.  Since quantum corrections to the RG flow are
small in this regime, the effective action should be \rwa{classically} scale invariant to
first approximation, with small quantum corrections.  In quantum terms, classical scale invariance is
broken by operators appearing with coefficients proportional to positive powers the cutoff.  This effective action has a smooth limit as $\L$ is taken to zero, since there are 
no long-range degrees of freedom to generate infrared singularities.

\subsection{Coupling to another sector}\label{sec:BTheory}

\heading{Dynamical source for the $a$-field}

Now suppose further that the vev of $a$ is not fixed arbitrarily, but set by a source term
$B(x) a\sqd$, in which case the vacuum expectation value of $a$ goes as $B\uu{+{1\over 4}}$.
For $\L\muchlessthan \L\ll{\rm mass}  \sim B\uu{+\hh}$, the effective action is classically scale-invariant as a functional of $B$ and its derivatives, with $B$ assigned a scaling dimension of $2$, and
quantum corrections suppressed by positive powers of ${\L\over{\sqrt{B}}}$.

We can go further and let $B(x)$ itself be a dynamical object, constructed as an operator in a
local effective field theory coupled to the $a$ sector.  This theory must also be regularized and
renormalized in its own right.  However if the sector in which the $B$ operator lives
also has trivial infrared dynamics -- either because it is gapped or because it becomes
free and massless at low energies -- then once again, the effective Lagrangian will have
only positive powers of $\L$ at the lowest energies.

\heading{Integrating out the $a$ field}

For purposes of examining physics of the $B$ sector at scales below $\L\ll{\rm mass}$, we can go further still and integrate out the $a$ field altogether.  We are then left with some effective theory
of the $B$ sector.  Here, powers of $B$ can appear in the denominators of
terms in the
effective theory, because $B$ sets the mass scale for the $a$ field.  However the derivatives
of $B$ only appear as polynomials in the numerator, in the expansion where $B$ is
varying slowly compared to the length scale $|B|\uu{-\hh}$ itself.

If the coupled dynamics of $a$ and $B$ are conformal below the UV scale $g\sqd$, then
the effective action in the $B$-sector must itself be invariant under the conformal symmetry 
at scales below the mass of $a$.  
This theory is under control when the conformal symmetry is broken spontaneously, since $\sqrt{B}$ will
always be comparable to the mass of $a$.   The effective action will contain terms such as
$|B|\uu {+{3\over 2}}, {{(\pp B)\sqd}\over{  |B|\uu{{3\over 2}}}},  {{(\pp\sqd B)\sqd}\over{  |B|\uu{{5\over 2}}}}$, et cetera. 

\heading{Quantum mechanics of the effective $B$-theory}
We can then quantize this effective theory in an expansion where the rate of variation of $B$ is
small compared to the cutoff $\Lambda$, which in turn is small compared to the scale 
$\sqrt{B} \sim \L\ll{\rm mass} \sim a\sqd$:
\bbb
{{ |\pp B|}\over B} \muchlessthan 
\L \muchlessthan \sqrt{B}\ .
\een{FirstDoubleHierarchy}

In this expansion, quantum corrections to
amplitudes are suppressed by $p / \sqrt{B}$ and $\L / \sqrt{B}$.  As a result,
the effective theory of the $B$ sector is conformally invariant, not only quantum
mechanically, but approximately classically as well.  Corrections to classical scale invariance
in the effective Lagrangian are given by terms with positive powers of ${\Lambda / \sqrt{B}}$.
In the case where the full theory is conformally invariant, the quantum mechanical, cutoff-dependent terms can be determined systematically
in terms of the classical, cutoff-independent terms through the renormalization group equation.

\newcommand*{\Scale}[2][4]{\scalebox{#1}{\ensuremath{#2}}}%

Let the effective Lagrangian for the $B$ sector be a sum of terms 
\begin{equation}
{\cal L}\ls\L [B ] \equiv {\cal L}\upp{\rm cl} + {\cal L}\upp{\rm qu}\ls\L\ ,
\end{equation}
where the terms ${\cal L}\upp{\rm cl}$ are classically scale-invariant, and the terms
${\cal L}\upp{\rm qu}\ls\L$ are not.  That is,\footnote{In fact, we can allow a slightly more general form of the expansion, with logarithms ${\tt ln} (B / \L\sqd)$ on top of powers of $\L$.  The logarithms are
physically quite interesting, but somewhat beside the point of the discussion here, so we do not emphasize them.  The only potentially important logarithm would be a term in the expansion
with cutoff dependence $\L\uu 0\cc ({\tt ln}(B / \L\sqd))\uu k$ for $k >0$.  But we have assumed the theory of the $B$ sector to be trivial as $\L\to 0$, which does not allow for such terms.}
\begin{equation}
{\cal L}\upp{\rm qu}\ls\L =\sum\ll{\D\ll c \neq 3} \cc\cc \L\uu{3-\D\ll{\rm c  } } \cc {\cal L}\upp{{\rm qu}}\ll{\D\ll c}\ ,
\end{equation}
where all terms in $ {\cal L}\upp{\rm cl} $ have classical scaling dimension $\D\ll{\rm c}$ equal to $3$,
and ${\cal L}\upp{{\rm qu}}\ll{\D\ll {\rm c}}$ has classical scaling dimension equal to $\D\ll {\rm c}$.

If ${{\d\uprm{RG}}\over{\d \L}} \cc {\cal L}$ is the renormalization of the Lagrangian as we integrate out a shell of modes of infinitesimal thickness $\d\L$, then the  equation for the theory to live at a 
fixed point is simply
\begin{equation}
 \L\cc {{\d\uprm{RG}}\over{\d \L}} \cc {\cal L} = \sum\ll{\D\ll c \neq 3} \cc\cc (\D\ll c - 3 ) \cc  \L\uu{3 - \D\ll{\rm c} } \cc {\cal L}\upp{{\rm qu}}\ll{\D\ll c}\ .
\end{equation}
Now, we are working in the limit \rr{FirstDoubleHierarchy}, so can expand both sides in powers of 
${\L\over{\sqrt{B}}}$.  By our assumption, the Lagrangian for the $B$ sector is trivial in the
infrared (either gapped or else free and massless), so the left-hand side has only terms with $\D\ll {\rm c}  < 3$.

This does \emph{not} mean, however, that the right-hand side contains only a finite number of
terms.  Since we are working in an effective theory where the operators are allowed to contain
powers of $B$ in the denominator, the classical scaling dimensions of the operators appearing in the $\L$-dependent terms can be \emph{negative}: The set of allowed values of $\D\ll{{\rm c}}$ is bounded
above by $+3$, but unbounded below.   

For instance, the RG flow of the classical piece of the Lagrangian may take the form
\begin{equation}
\L{{\d\uprm{RG}}\over{\d \L}} \cc {\cal L}\upp{cl} = K\ll 1 \cc \L\uu 3 + K\ll 2 \cc {{\L\uu 5}\over B}
+ K\ll 3 \cc \L\uu 5 \cc {{(\pp B)\sqd}\over{B\uu 4}} + \cdots
\end{equation}
Then the leading pieces of the quantum action must be
\begin{equation}
{\cal L}\upp{\rm qu}\ls\L = - {1\over 3} \cc K\ll 1 \cc \L\uu 3 - {1\over 5} K\ll 2 \cc {{\L\uu 5}\over B}
- {1\over 5} K\ll 3 \cc \L\uu 5 \cc {{(\pp B)\sqd}\over{B\uu 4}} + \cdots
\end{equation}

Thus we can "bootstrap" the coefficients of the operators in the $\L$-dependent quantum terms ${\cal L}\uprm{qu}$ algorithmically from the form of the classically scale-invariant Lagrangian ${\cal L}\uprm{cl}$, though ${\cal L}\uprm{cl}$ may itself have unknown coefficients.

\subsection{Lessons for large-$J$ perturbativity}

Though the discussion at the level of generality in this section may be unfamiliar,
many examples of this type of dynamics are already well known.
The most familiar examples occur in supersymmetric
theories with moduli spaces of exactly supersymmetric vacua.  In these theories, the
role of the $B$ sector is played by the massless moduli fields and unbroken gauge fields themselves, while the role
of the $a$ sector is played by the massive, spontaneously broken gauge fields and the 
scalars that gain a potential through $F$-- and $D$--terms.  

The toy model above contains the essence of the large-$J$ analysis of the critical $O(2)$ model in three dimensions and that of its supersymmetric cousin, the superconformal $W = \Phi\uu 3$ model.  In these theories, the role of the $a$ field is played by the magnitude of the complex scalar $\phi$, and the $B$ operator is
constructed from the gradient squared of the phase variable $\chi$, that is, $B(x) \equiv |\pp\chi|\sqd$,
which is proportional to the charge density of the system.

Our large-$J$ descriptions of the conformal $O(2)$ model and $W = \Phi\uu 3$ model will be effective theories for the phase variable $\chi$ after integrating out the 
magnitude $a \equiv |\phi|$ of the complex scalar.  In the supersymmetric $W = \Phi\uu 3$ 
theory, the fermions will also acquire a gap of order $m\ll a \sim \sqrt{B} = |\pp\chi|$, and be
integrated out, leaving us with the effective theory of $\chi$ alone.

Keeping in mind the toy example of this section will guide us through the theory-dependent details of the renormalisation-group analysis of the critical $O(2)$ model and superconformal $W=\Phi\uu 3$ model.

\section{Bosonic $O(2)$ model in 3D}
\label{sec:O2-3D}

\subsection{$O(2)$ model and its large-$J$ analysis}\label{sec:o2}

Now we apply the lessons of the toy model to the full $O(2)$ model.  The UV Lagrangian of the system, which
is stable and renormalizable, is of the form
\begin{equation}
\mathcal{L}_{\text{UV}}=-\partial_\mu\varphi^*\partial^{\mu}\varphi-c|\varphi|^2-g^2|\varphi|^4,
\end{equation}
where $\varphi$ is a complex scalar and $c$ is fine-tuned in order for the Lagrangian
to flow to a conformal IR fixed point~\cite{Wilson:1971dc}:
\begin{equation}
\text{UV theory}\xrightarrow{\text{RG flow}}\text{IR conformal fixed point}.
\end{equation}
Also in the IR, our degree of freedom is a complex scalar $\phi$.
In order for us to make conceptual contact with the toy problem of Section~\ref{sec:BigJRG},
we parametrize $\phi$ as
\begin{equation}
  \phi=a\, e^{i\chi},
  \label{PolVars}
\end{equation}
where $a\in\mathbb{R}^+$ and $\chi$ is $2\pi$--periodic.

The full exact Wilsonian RG equations are complicated.  They simplify quite a bit in the large-$J$ limit, but not until the cutoff is lowered to $\sqrt{\r}$.  The charge density $\r$ cannot affect the flow between the scale $\L\ll{\rm UV} \equiv g\sqd$
and the scale $\L = \sqrt{\rho}$. Between these scales, the renormalisation group flow is unaffected by the vev.
In this range, the equations contain all the complexity of the full Landau--Ginzburg theory in its strongly coupled regime, including its flow to the conformal fixed point.  This range of the RG flow can be treated only numerically in a standard treatment, 
and is not simplified by large $J$ at all.  It is only when we reach the scale $\L = \sqrt{\rho}$ that the RG equations simplify.

We find, however, that if we \rwa{use as input} the fact that this theory flows to a fixed point, we
can strongly constrain the form of its Wilsonian 
effective action in the regime of large charge density.  When $\r\uu{+\hh}$ is
much larger than the cutoff, the magnitude field $a$ decouples as in the toy model
of the last section.  Let us now study the effective action for the complex scalar 
in the regime 
\bbb
\L \muchlessthan a\sqd \muchlessthan g\sqd.
\een{DoubleHierarchyB}

\heading{Approximate Classical Scale Invariance}

In this limit we write the Lagrangian as\footnote{The normalization condition for the field has been chosen so that the 
kinetic term of $a$ is unit and canonical.}
\begin{equation}
\mathcal{L}_{\text{IR}}=-\frac{1}{2}(\partial_\mu a)^2
-f(a)(\partial_\mu \chi)^2-V(a)+(\text{higher derivative terms}).
\label{IRLagForm}
\end{equation}

At the IR fixed point, the Lagrangian is approximately classically scale-invariant, with
corrections to classical scale invariance that go as $\L / a\sqd$. The leading Lagrangian
density must have mass dimension \(3\), and the fields have dimensions
\begin{align}
  a& \propto[\text{mass}]^{1/2}, &
                                   \chi &\propto[\text{mass}]^{0}.
\end{align}
The former is fixed by the normalization convention for the kinetic term of $a$, and the latter
is fixed by the dimensionless periodic identification of $\chi$.

It follows that the functions in the Lagrangian have to scale as
\begin{align}
  V(a) &\propto a^6, & f(a)&\propto a^2.
\end{align}
Therefore the Lagrangian at the IR fixed point can be written as
\begin{multline}
  \mathcal{L}_{\text{IR}}=- \frac{1}{2}(\partial_\mu a)^2
  -\frac{1}{2}\kappa a^2 (\partial_\mu \chi)^2-{{h\sqd}\over{12}} a^6 +
(\text{Ricci coupling}) \\+(\text{higher derivative terms}),
  \label{IRFormWitha}
\end{multline}
where \(\kappa \) and \(h\sqd\) are numerical constants.

We solve the Euler--Lagrange equation using the fact that the lowest-energy solution at fixed and large charge is homogeneous in space (no dependence on the spatial coordinates) and find that the equilibrium value of $a$ lies at 
\bbb
a\uu 4 = - {{2\k}\over{h\sqd}}\cc (\pp\chi) \sqd =  + {{2\k}\over{h\sqd}}\cc |\pp\chi|\sqd\ .
\een{aFromPartialChi}
(Here, we have used the fact that the gradient of $\chi$ is timelike around a state of
large charge density.)
The charge density $\r = \delta \mathcal{L}_{\text{IR}} / \delta \dot \chi = \k \cc a\sqd \cc \dot{\chi}$ is then
\begin{equation}
\r = \sqrt{{{\k h\sqd}\over 2}} \cc a\uu 4\ ,
\end{equation}
and the total charge is
\begin{equation}
  J = \int_{S^2} d^2 x \, \rho = 4 \pi R^2 \sqrt{\frac{\kappa h^2}{2} } a^4 .
\end{equation}
In terms of the charge density $\r$,
our double hierarchy \rr{DoubleHierarchyB} becomes
\bbb
\L \muchlessthan \sqrt{\r} \muchlessthan g\sqd.
\een{DoubleHierarchyC}

Since we are working on a sphere of radius \(R\), in order for the effective action to be consistent, \(\Lambda\) must be parametrically larger than the IR cutoff \(M_{\textsc{ir}} = 1/ R\). This means that \(\Lambda\) is in the regime
\begin{equation}
  \frac{1}{R} \muchlessthan \Lambda \muchlessthan \frac{\sqrt{J}}{2 R \sqrt{\pi}} \muchlessthan g^2 
\end{equation}
which can only be consistent if the charge is large:
\begin{equation}
  J \muchgreaterthan 1 .  
\end{equation}

\heading{Renormalization Group Analysis}
Let us now analyze the RG equations for the effective action \rr{IRLagForm}, using
the property that the underlying theory is conformally invariant.  In particular, let us show
that the corrections to classical scale invariance are small, and that the cutoff-dependent
quantum terms can be derived algorithmically from the cutoff-independent classical terms
by the renormalization group equation.

Now let us prove that the classical scaling above is correct for large enough charge density $\r$.
For this purpose, we want to estimate the change of $f(a)$ and $V(a)$ in \rr{IRLagForm} 
with respect to the change in
the momentum cut-off $\Lambda$.

We first Taylor expand around the minimum of $a$, which will be denoted $a_0$.
We also let $\hat{a}=a-a_0$:
\begin{equation}
  \begin{split}
    \mathcal{L}_{\text{IR}} ={}& -\frac{1}{2}(\partial_\mu \hat{a})^2 \\
    &-V(a_0)-\hat{a}V^{\prime}(a_0)-\frac{1}{2}\hat{a}
    V^{\prime\prime}(a_0)+\cdots \\
    &-f(a_0)(\partial_\mu \chi)^2-\hat{a}f^{\prime}(a_0)(\partial_\mu
    \chi)^2
    -\frac{1}{2}\hat{a}^2f^{\prime\prime}(a_0)(\partial_\mu \chi)^2+\cdots \\
    &+(\text{higher derivative terms}).
  \end{split}
\end{equation}
Here the Feynman rules are as follows
:
\begin{eqnarray}
\parbox[c]{31mm}{\includegraphics[width=3cm]{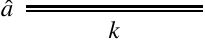}}&=&\frac{1}{k^2+V^{\prime\prime}(a_0)}
\\
\parbox[c]{31mm}{
\includegraphics[width=3cm]{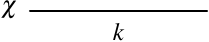}
}
&=&\frac{1}{2f(a_0)k^2}\\
\parbox[c]{31mm}{
\includegraphics[width=3cm]{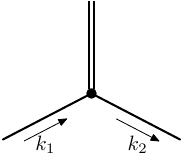}
}
&=&f^{\prime}(a_0)(k_1\cdot k_2)\\
\parbox[c]{31mm}{
\includegraphics[width=3cm]{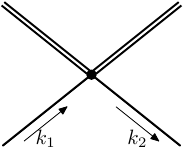}
}
&=&\frac{1}{2}f^{\prime\prime}(a_0)(k_1\cdot k_2)\\
\end{eqnarray}
The double lines represent $\hat{a}$-propagators and the single lines represent
$\chi$ propagators.
Apart from the ones shown above, we have one more diagram representing the source term
$\hat{a}V^{\prime}(a_0)$:
\begin{eqnarray}
\parbox[c]{31mm}{
\includegraphics[width=3cm]{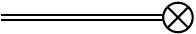}
}
&=&V^{\prime}(a_0)
\end{eqnarray}

First we renormalize the one-point function. As we have defined
the effective potential is at its minimum when $\hat{a}=0$, 
the full one-point function becomes zero. Therefore, we shall not worry about
one-point functions anymore.

Diagrams that contribute to
the renormalization of $f(a)$ at 1-loop are 
\begin{eqnarray}
&\text{(a)}&
\parbox[c]{41mm}{
\includegraphics[width=4cm]{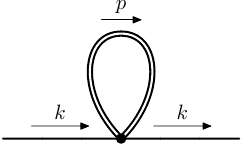}
}\\[1cm]
&\text{(b)}&
\parbox[c]{41mm}{
\includegraphics[width=4cm]{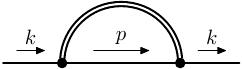}
}
\end{eqnarray}
We now calculate these diagrams. We will take the momentum cut-off $\Lambda$
and integrate the loop momentum over a shell with radius $\Lambda$ and thickness 
$\delta\Lambda$.
\begin{itemize}
\item[(a)]The contribution from diagram (a), $\delta_1f(a_0)$, is given by
\begin{equation}
  \begin{split}
    \delta_1 f(a_0)&\propto\frac{1}{k^2}\int\limits
    ^{\Lambda+\delta\Lambda}_\Lambda\!
    d^3p\,\frac{k^2f^{\prime\prime}(a_0)}{p^2+V^{\prime\prime}(a_0)}\\
    &\sim \frac{f^{\prime\prime}(a_0)\Lambda^2\delta\Lambda}
    {\Lambda^2+V^{\prime\prime}(a_0)}\sim
    \frac{f^{\prime\prime}(a_0)\Lambda^2\delta\Lambda}
    {V^{\prime\prime}(a_0)}.
  \end{split}
\end{equation}
The last approximation holds when $\r / \L\sqd$ is large enough: $a_0\propto \rho^{1/4}$
becomes large enough to satisfy $V^{\prime\prime}(a_0)\muchgreaterthan \Lambda^2$
in our limit $\r \muchgreaterthan \L\sqd$.
\item[(b)]The contribution from diagram (b), $\delta_2f(a_0)$, is given by
\begin{equation}
  \begin{split}
    \delta_2 f(a_0) &\propto \frac{1}{k^2}\int\limits
    ^{\Lambda+\delta\Lambda}_\Lambda \!d^3p\,
    \frac{1}{(k-p)^2+V^{\prime\prime}(a_0)}
    \frac{\left(f^{\prime}(a_0)\right)^2(k\cdot p)^2}{f(a_0)p^2}
    \\
    &\sim \int\limits ^{\Lambda+\delta\Lambda}_\Lambda
    \!d|p||p|^2\!\!\!\int\limits^{\cos(\theta)=1}_{\cos(\theta)=-1}
    \!d(\cos(\theta))
    \frac{(f^{\prime}(a_0))^2)\Lambda^2\cos^2(\theta)}
    {V^{\prime\prime}(a_0)f(a_0)\Lambda^2}\\
    &\sim
    \frac{(f^{\prime}(a_0))^2\Lambda^2\delta\Lambda}{V^{\prime\prime}(a_0)
      f(a_0)}.
  \end{split}
\end{equation}
Again the approximation holds when $\r$ is large enough compared to $\L$.
\end{itemize}

Summing over the two contributions above, 
we find the RG equation below for the quantum scaling (assuming that $f(a)\propto a^n$):
\begin{equation}
\Lambda\left.
\frac{\delta}{\delta\Lambda}\right|_\text{q}
f(a_0)\propto\frac{f(a_0)}{a_0^2}
\frac{\Lambda^3}
{V^{\prime\prime}(a_0)}.
\label{qu}
\end{equation}
We can compare this with its classical scaling, which is of course just
proportional to $f(a\ll 0)$ itself.
Therefore the ratio of the quantum scaling over the classical one is
\begin{equation}
\frac{\Lambda\left.
\frac{\delta}{\delta\Lambda}\right|_\text{q}
f(a_0)}{
f(a_0)
}=\left(\frac{\Lambda}{a^2}\right)^3.
\end{equation}
This is small when $\r$ is large enough, \emph{i.e.} when $a^2\muchgreaterthan\Lambda$.
The analysis for $V(a_0)$ goes parallel to the analysis above:
\begin{equation}
\frac{\Lambda\left.
\frac{\delta}{\delta\Lambda}\right|_\text{q}
V(a_0)}{
V(a_0)
}=\left(\frac{\Lambda}{a^2}\right)^3.
\end{equation}
Just as in the discussion in Section~\ref{sec:BTheory}, we can derive
all $\L$-dependent coefficients in the effective Lagrangian from the classically scale-invariant
ones, through the fixed point condition in the renormalization group equation.

In the limit $\L\to 0$ with the field configuration and in particular $a \neq 0$  held fixed, the 
Wilsonian action is finite.  This is clear, because the only possible singularities as $\L\to 0$
would come from integrals at low momentum.  Since $a$ is gapped and $\chi$ is infrared-free in the absence of interactions with $a$, there are no diagrams that can generate
such singularities.

\subsection{Reduction to Goldstones}\label{sec:red-goldstone}

\heading{Classical elimination of $a$}

Starting with \rr{IRFormWitha}, we now integrate out the $a$ field and examine the effective action for the $\chi$ field.  The mass term of the $a$ field is proportional to $|\pp \chi|=\sqrt{-\del_\mu \phi \del^\mu\phi}$, so
in the limit $|\pp\chi| \muchgreaterthan \L$ we can set the $a$ field classically to its
equilibrium value \rr{aFromPartialChi}, with
the quantum corrections from its fluctuations being proportional to positive powers of
${\L\over{|\pp\chi|}}$.  Using \rr{IRFormWitha} we have
\bbb
{\cal L} = b\ll\chi \cc |\pp\chi|\uu 3 +
(\text{lower order in $|\pp\chi| $})\ ,  
\een{LeadingGoldstoneAction}
where 
  $b\ll\chi \equiv {{\sqrt{2}}\over 3} \cc {{\k\uu {3\over 2}}\over h}$.

\heading{Terms of lower order in $\pp\chi$}

The higher-derivative terms come from the kinetic term for $a$, as well as from higher-derivative
terms in the effective action~\rr{IRFormWitha}.  The kinetic term for $a$ generates higher-derivative terms such as ${{(\pp|\pp\chi|)\sqd}\over{|\pp\chi|}}$.  This term comes along with a
coupling to the Ricci scalar, proportional to ${\tt Ric}\ll 3 \cc |\pp\chi|$.  The coefficients of
these two terms are not independent; Weyl invariance relates them to one another.

There is an infinite series of such terms in the effective action, with arbitrary derivatives of $\chi$ in the numerator, and only $|\pp\chi|$ occurring to negative powers.  These operators are classically scale-invariant, and are arranged hierarchically in terms of the number of powers of $\pp\chi$ that occur in them.  This is the natural organization of operators when we compute observables in a state of large and approximately constant density,
\begin{equation}
{{\pp\r}\over \r} \muchlessthan \L \muchlessthan \sqrt{\r}\ .
\end{equation}
In Subsection \ref{sec:class-goldstones}, we will classify the first
few leading operators in this expansion of ${\cal L}$.\footnote{This
  is in the spirit of the functional RG flow (see~\cite{Kopietz:2010zz} and references therein). Note that our analysis is novel because we are not making approximations such as large \(N\) or \(\epsilon\) expansion, but we hold the dimensionality of space and the field content constant.}  First, we briefly analyze the quantum properties of the conformal Goldstone
theory.

\subsection{Conformal Goldstones at the quantum level}\label{sec:QuantumGoldstone}
\label{sec:QuantumGoldstones}

At first sight, the action~\rr{LeadingGoldstoneAction} looks quite strange.  The form of the Lagrangian density is singular and infinitely strongly coupled as an action expanded around the origin $\chi = 0$.  But
the action~\rr{LeadingGoldstoneAction} is not meant to be used there: It is only a first term in
an infinite series of terms with higher powers of $|\pp\chi|$ in the denominator.  

Rather, the action is meant to be expanded around a background $\pp\chi\ll 0 \propto \sqrt{\r\ll 0}$ of constant charge density, and the field $\chi$ is quantized in small fluctuations whose wavelength is long compared to $\r\ll 0\uu{-\hh}$.  In this limit, the Lagrangian is fully under
perturbative control.

To see this, break up $\chi$ into a background and fluctuation, $\chi = \chi\ll 0 + \chi\ll{\rm fluc}$,
where $\chi$ has constant gradient. 
 The expansion around a fixed background $\chi\ll 0$ takes the form\footnote{We have dropped the term linear in the fluctuations, which is a total derivative. This term 
 is important in that it contributes to the form of the expression for the conjugate momentum to $\chi$ and therefore
 the charge density, in the small-fluctuation expansion.  However it does not
 affect the equations of motion or the computation of corrections, so we omit it in the present discussion.}
 \begin{equation}
 {\cal L} = b\ll\chi \cc |\pp\chi\ll 0|\uu 3 + |\pp\chi\ll 0 | \cc \big [ \cc {\rm quadratic~in~}\pp \chi\ll{\rm fluc} \cc 
 \big ] +  [ \cc {\rm cubic~in~}\pp \chi\ll{\rm fluc} \cc 
 \big ]  + O \big ( {{(\pp\chi\ll{\rm fluc})\uu 4}\over{|\pp\chi\ll 0|}} \big ) \ ,
 \end{equation}

 So the quantum mechanical fluctuations of $\chi\ll{\rm fluc}$ are of order $|\pp\chi\ll 0|\uu{-\hh}$,
 or equivalently order $\r\uu{-{1\over 4}}$.  It is sometimes convenient to work in terms
 of a scalar field $\hat{\chi} \equiv |\pp\chi\ll 0|\uu{+\hh} \cc\chi\ll{{\rm fluc}}$ with unit-normalized kinetic term.  In terms
 of the hatted Goldstone, we have
 \bbb
  {\cal L} = b\ll\chi \cc |\pp\chi\ll 0|\uu 3 + \big [ \cc {\rm quadratic~in~}\pp\chh \cc 
 \big ] + |\pp\chi\ll 0|\uu {-{3\over 2}} \cc  [ \cc {\rm cubic~in~}\pp\chh \cc 
 \big ]  + O \big ( {{(\pp\chh)\uu 4}\over{|\pp\chi\ll 0|\uu 3}} \big ) .
 \een{CanonicalizedGoldstoneAction}
The quantum fluctuations of the canonically normalized field $\chh$ are of order $1$,
and all the $\r$-dependence comes through coefficients of the vertices which are
powers of $\pp\chi\ll 0 = O(\r\uu{+\hh})$.  The
effective coupling constant that suppresses each successive quantum loop is therefore 
$|\pp\chi\ll 0|\uu{-3} \propto \r\uu{-{3\over 2}}$.  

There is a simple rule to keep track of the quantum-mechanical $\r$-scaling of a given term,
without explicitly breaking up the $\chi$ field into background and fluctuation.  Since the background
$\chi\ll 0$ has constant gradient, then the term $\pp\chi$ is dominated by the background piece $\pp\chi\ll 0$, while the terms $\pp\uu k\chi$ for $k\geq 2$ are purely fluctuation pieces, since
$\pp\uu k\chi\ll 0 = 0$.  So the rules are
\begin{align}
\pp\chi \sim \pp\chi\ll 0 &= O(\r\uu{+\hh})\ , 
\label{RhoCountingRulesA}
\\
\pp\uu k \chi = \pp\uu k\chi\ll{{\rm fluc}} = |\pp\chi\ll 0|\uu{-\hh} \cc\pp\uu k \chh &= O(\r\uu{-{1\over 4}})\ .
\label{RhoCountingRulesB}
\end{align}
These rules will allow us to make quick work of the enumeration of the first sub-leading operators
in the Goldstone effective action, in Section~\ref{sec:OpClass}.

\heading{Cutoff-dependent terms}

The effective Lagrangian for $\chi$ is classically scale-invariant in the large-$\r$ limit, and
also fully quantum-mechanically scale invariant by virtue of our input assumption that
it is the effective theory of an exactly conformally invariant theory.  As a result, one
needs in general to include cutoff-dependent terms to restore conformal
invariance at the quantum mechanical level.

Just as in the case of the toy model of Section~\ref{sec:BTheory}, we can compute the
quantum terms algorithmically in a $\L\uu 3 / |\pp\chi|\uu 3$ expansion, from the starting
point of the $\L$-independent, classically scale-invariant terms in the Lagrangian.  The form of these can be seen by taking the cutoff-dependent terms
in the $\phi$ system and reducing those to Goldstone terms by integrating out $a = |\phi|$.  Alternately, a renormalization group equation for the $\chi$ action 
determines the cutoff-dependent terms directly, in terms of the canonically scale-invariant terms in the Goldstone action.  The quantum pieces will consist of terms such as $\L\uu 3, {{\L\uu 3 \cc (\pp\sqd\chi)\sqd}\over{|\pp \chi|\uu 4}}, \cdots$. 

The cutoff-dependent terms are of course scheme-dependent and not very interesting in
themselves, however
important they may be as a point of principle for establishing the self-consistency of our treatment of the theory at large $J$.  For low-order calculations, such as the one we will perform in Section~\ref{sec:large-J-op-dim}, an analytic renormalization scheme such as dimensional regularization
or $\z$-function regularization, is far more convenient, as it subtracts the counterterms
automatically and we never have to calculate the explicit form of the quantum terms in
the action.  In Section~\ref{sec:large-J-op-dim} we will indeed use $\z$-function regularization, and therefore we will not pursue the calculation of the cutoff-dependent terms further.

\subsection{Classification of operators in the conformal Goldstone action}\label{sec:class-goldstones}
\label{sec:OpClass}
Let us now classify all possible scale-invariant operators appearing in the conformal Goldstone action, according to their $\r$-scaling.   In order to understand the $\r$-scaling of a given operator at the quantum level, we use the simple $\r$-counting rules \rr{RhoCountingRulesA}, \rr{RhoCountingRulesB}
derived in Section~\ref{sec:QuantumGoldstones}.

We are going to write down scale-invariant operators at order $\r^{\alpha \geq 0}$.
The scalings to remember are
$\partial\chi\sim \r^{1/2}$ and 
$\partial\dots\partial\chi\sim \r^{-1/4}$.
  We retain only
scalar operators of scaling dimension $3$, including curvatures of the background metric in
our counting.

\heading{Order $\r\uu{+{3\over 2}}$}

There is only one operator of order  $\r\uu{+{3\over 2}}$: The leading-order Lagrangian 
\bbb
\co\ll{{3\over 2}} \equiv |\pp\chi|\uu 3\ .
\een{Op3Ov2}
This term is conformally invariant as well as scale-invariant, as an operator in the Lagrangian density.


\heading{Order $\r\uu{+{1\over 2}}$}

At this order, there are two operators and they are both curvature-dependent.
The first is
\bbb
\co\ll{{1\over 2}} \equiv {\tt Ric}\ll 3 \cc |\pp\chi|\ ,
\een{Op1Ov2}
 where by ${\tt Ric}\ll 3$ we mean the three-dimensional Ricci
scalar.  This term is scale-invariant as a term in the action
but \rwa{not} conformally invariant.  This is related to
the fact that the Ricci scalar transforms inhomogeneously under a Weyl transformation
of the background metric $g\ll{\bullet\bullet} \to \exp{2\upsilon(x)} \cc g\ll{\bullet\bullet}$.
To restore Weyl-invariance and conformal invariance, this term must be completed
with an operator of the form ${{(\pp|\pp\chi|)\sqd}\over{|\pp\chi|}}$ in the Weyl-invariant combination
\bbb
\hat{\co}\ll{{1\over 2}} \equiv {\tt Ric}\ll 3 \cc |\pp\chi| + 2 \cc {{(\pp|\pp\chi|)\sqd}\over{|\pp\chi|}}\ .
\een{Op1Ov2WeylCompleted}
Note that the $\r$-scaling of the completing term is negative.

There is also the operator
\begin{equation}
{{R\ll{\m\n} (\pp\uu\m\chi)(\pp\uu\n\chi)}\over{|\pp\chi|}}\ .
\end{equation}
This operator is of order $\r\uu{+\hh}$ for a generic background metric.  However we will be mostly interested in the case of an unwarped product metric of the form $({\tt time}) \times ({\tt spatial~slice})$, and a spatially homogeneous solution $\chi\ll 0$.  For such a metric, the only non-vanishing components of $R\ll{\m\n}$ are the spatial components, which have vanishing
contraction with the background gradient $\pp\ll\m\chi\ll 0$.  Therefore this operator contributes only through fluctuation terms.  Replacing each of the two contracted $\pp\chi\ll 0$'s with a
$\pp\chi\ll{{\rm fluc}} = |\pp\chi\ll 0|\uu{-\hh} \cc \pp\chh$,
the scaling of the operator comes down to $\r\uu{-1}$, which is below $0$ and so we
discard it.

\heading{Order $\r\uu{+{1\over 4}}$}

We obtain two possible terms at order $\r\uu{+{1\over 4}}$:
\begin{eqnarray}
|\partial\chi|\partial_\mu\partial^\mu\chi&\sim&\r^{1/4}\label{kisa},\\
\partial_\mu\chi\partial^\mu|\partial\chi|=\frac{\partial_\mu\partial_\nu\chi
\partial^\nu\chi}{|\partial\chi|} &\sim&\r^{1/4}\label{wei}.
\end{eqnarray}

These terms are not linearly independent as operators; one combination vanishes by
virtue of the equations of motion.  The remaining linear combination is non-vanishing as 
an operator, but does not appear in the effective Lagrangian, because it is odd under
the charge-conjugation symmetry of the theory:
\bbb
\phi\leftrightarrow \phb, \llsk\llsk \chi \to - \chi\ .
\een{ChargeConj}
Hence there are no terms of order $\r\uu{+{1\over 4}}$ appearing in the Lagrangian.

\heading{Order $\r\uu 0$}

There are no operators at order $\r\uu 0$.  This is a key fact so let us establish it formally.
Since first derivatives of $\chi$ scale as $\r\uu{+\hh}$ and higher-derivatives of 
$\chi$ scale as $\r\uu{-{1\over 4}}$, the number of $\chi$'s with higher derivatives on them
must be even for an operator to scale as a half-integer power of $\r$.  Dimension-3 operators with $\pp\chi$'s only are of the form $(\pp \chi)\uu {3-k} {\cal G}\ll k$, where ${\cal G}\ll k$
is a geometric invariant of scaling dimension $k$,
constructed from background curvatures.  For $k= 3$ there 
are no such operators.\footnote{The gravitational Chern--Simons term, which has dimension $3$, is not gauge-invariant when multiplied by other operators such as $\pp\chi$.}

The next possibility would be terms with two or more $\chi$'s in the numerator with higher derivatives acting on them.  The total dimension of the numerator would be at least four, and have $\r$ scaling bounded above by $\r\uu{-\hh}$.  The dimension of the operator could then only be brought down
to $3$ by putting one or more factors of $|\pp\chi|$ in the denominator, which would drive the
total $\r$-scaling of the operator to $\r\uu{-1}$ or less.  

We conclude there are no operators of order $\r\uu 0$.  This fact has physical significance, and we shall see in Section~\ref{sec:observables} that the $J\uu 0$ term in the expansion of
the operator dimension at large $J$, is calculable and universal as a result.
\def\OPDEFOUT{ 

\heading{Operators of the form $\partial\partial\partial\partial\partial\chi\chi\chi\chi/|\partial\chi|^2$
and higher terms}
We cannot produce rotationally-invariant operators of the form
$\partial\partial\partial\partial\partial\chi\chi\chi\chi/|\partial\chi|^2$
that does not include $|\partial\chi|$ or $\partial|\partial\chi|$:
we do not have any operators of this form which have not or will not be counted
as operators of other forms.
As to terms of the form
 $\partial\partial\partial\partial\partial\partial\chi\chi\chi\chi\chi/|\partial\chi|^3$
and other higher terms, again we cannot produce any operators
that have not been counted already. This is because we cannot avoid having
$\partial_\mu\chi\partial^\mu$ in the numerator, which then will be cancelled
by $|\partial\chi|^2$ in the denominator.

\subsubsection{Operators of the form {$\partial\partial\partial\chi$}}

\heading{Operators of the form $\partial\partial\partial\chi/1$}

\shg{We might not have to discuss these operators at all.  We can use an independent analysis of curvature dependent terms and not worry at first about the Weyl-completion.}

The only possibility of the term of this form is thus:
\begin{equation}
\partial_\mu\partial^\mu|\partial\chi|.
\end{equation}
This again can be calculated explicitly;
\begin{equation}
\partial_\mu\partial^\mu|\partial\chi| = 
\frac{\partial_\mu\partial^\mu\partial_\nu\chi\partial^\nu\chi}{|\partial\chi|}
+\frac{\partial_\nu\partial^\mu\chi\partial^\nu\partial_\mu\chi}{|\partial\chi|}
-
\frac{\partial^\nu\partial_\mu\chi\partial_\mu\partial_\alpha\chi\partial^\nu
\chi\partial^\alpha\chi}{|\partial\chi|^3}.
\end{equation}

Those terms will be counted later, so we will not be bothered by those terms for now.
\heading{Operators of the form $\partial\partial\partial\partial\chi\chi/|\partial\chi|$}
There are three terms of this form:
\begin{eqnarray}
\frac{\partial_\mu\partial^\mu\partial_\nu\chi\partial^\nu\chi}
{|\partial\chi|}&\sim&J^{-1/4}<J^0\\
\frac{\partial^\mu\partial^\nu\chi\partial_\mu\partial_\nu\chi}
{|\partial\chi|}&\sim&J^{-1}<J^0\\
\frac{\partial^\mu\partial_\mu\chi\partial^\nu\partial_\nu\chi}
{|\partial\chi|}&\sim&J^{-1}<J^0.
\end{eqnarray}
They all $J$-scale sub-zero.

\heading{Classically contributing and noncontributing bosonic operators}\label{NonContrib}

There may be a few operators with positive $J$-scaling.  But they do \rwa{not} contribute to the energy of the helical classical solution on $S\uu 2$.  There is, for instance, the curvature-dependent operator $R\ll{\m\n}(\pp\uu\m \chi)
(\pp\uu\n \chi)$. 

 \shg{That's really the only relevant example I can think of here.  But maybe it's important to think through the principle behind the non contribution of these operators.
We should have a less ad hoc way of dealing with them and demonstrating their non contribution.}

} 

\section{The Supersymmetric $W = \Phi^3$ Model}\label{sec:phi3}

\subsection{The model}

Consider the ${\cal N}=2$ supersymmetric theory in three space-time dimensions, with a single chiral superfield, K\"ahler potential $K = \Phi\dag\Phi$ and superpotential $W = {1\over 3} \Phi\uu 3$.  
This theory is known to flow to an interacting superconformal fixed point, as can be shown, \emph{e.g.}, by the techniques of~\cite{Barnes:2005bm,Jafferis:2010un}.
 The $R$-charge
and dimension of $\Phi$ at the fixed point are equal to $+{2\over 3}$.  This theory has no marginal deformations and no small parameter of any kind.  Nonetheless we would like to analyze
its spectrum of operator dimensions.  It is equipped with a continuous global symmetry, namely the $R$-symmetry itself, and one can inquire again about the dimension of the lowest operator $\kket J$ of $\phi$-charge $J$.  (We define the charge
here to be ${3\over 2}$ times the $R$-charge, so that $ \phi$ has $\phi$-charge $+1$ and the supercurrent $Q\ll\a$ has $\phi$-charge $-{3\over 2}$.  When referring to the "charge" without
specifying, we will mean the charge normalised as $\phi$-charge rather than $R$-charge.)

\subsection{RG flow in the $W = \Phi\uu 3$ model}\label{sec:phi3-large-J}

Following both our toy model and the example of the \(O(2)\) model in Section~\ref{sec:O2-3D}, we expect large-$J$ perturbativity to mean that the flow of the Lagrangian with ${\tt log}(\L)$ is parametrically suppressed relative
to the original Lagrangian.  This is indeed the case; the calculation is mostly parallel to that case and so we suppress most details.

In our normalization convention for the fields, $W$ is taken to be unit-normalized (and thus has scaling \([\text{mass}]^2\)).  This is different from the standard convention in which the kinetic term is unit-normalized, but it is a far more convenient convention for a supersymmetric theory with superpotential, due to the non-renormalization theorem for $F$-terms. Since the superpotential is
\begin{equation}
  W = \frac{\Phi^3}{3}, 
\end{equation}
it follows that \(\Phi\) has dimensions \(\Phi \propto [\text{mass}]^{2/3}\). 
The Lagrangian must be classically scale invariant: in three spacetime dimensions it follows that the K\"ahler potential has dimension $1$ and it must scale as \(K \propto \abs{\Phi}^{3/2}\).
We fix the proportionality constant to
\begin{equation}
  K = \frac{16 b_K}{9} \abs{\Phi}^{3/2},
\end{equation}
which gives the kinetic term
\begin{equation}
  \mathcal{L}_{\text{kin}} = b_K \abs{\phi}^{-1/2} \vec \partial \phi \vec \partial \bar \phi ,
\end{equation}
and the potential
\bbb
V = {1\over{b\ll K}} |\phi|\uu{9\over 2}\ .
\een{SUSYPotential}
The Yukawa coupling
\bbb
{\cal L}\ll{\rm Yuk} =  i\cc \phi \cc \psi\ll \uparrow \psi\ll\downarrow + ({\rm h.c.})
\een{SUSYYukawa}
does not involve the K\"ahler potential at all.

\heading{Inclusion of the one-loop renormalization}

Now we apply the RG equations at large $|\phi|$ at the fixed point, including the small one-loop renormalization of the action.  (For the running of the K\"ahler potential, for instance, 
one can use~\cite{Brignole:2000kg}, generalizing an earlier formula~\cite{Grisaru:1979wc} in the renormalizable case.)

In three dimensions, we have:
\begin{equation}
  \L \cc \frac{dK}{d\Lambda} \propto \L \cc \ln \big [ \L\sqd + m\sqd \big ]\ ,
\end{equation}
where
\begin{equation}
m\sqd = \frac{|W\prpr|\sqd}{K\ll{\phi\phb}\sqd}  =  \frac{g\sqd \cc |\phi|\sqd }{ K\ll{\phi\phb}\sqd} .
\end{equation}
We expand the K\"ahler potential in powers of $|\phi|$ as
\begin{equation}
  K(\Phi) = K^{(0)}(\Phi) + K^{(1)} (\Phi) = \frac{16 b_K}{9} |\Phi|\uu{3\over 2} + K\upp 1 + \cdots .
\end{equation}
If we assume that the mass scale \(M = |\hat \phi|^{3/2}\) defined by the vev satisfies
\begin{equation}
  | \hat \phi |^{3/2} \muchgreaterthan \Lambda 
\end{equation}
and use the condition \(K^{(1)} \muchlessthan K^{(0)} \simeq |\hat \phi|^{3/2}\) we obtain a self-consistent expansion of the logarithm,
\begin{equation}
  \ln [\Lambda^2 + m^2] = \frac{\hat b_K \Lambda^2}{|\hat \phi|^3} - \frac{2 | \hat \phi|^{1/2}}{\hat b_K} K^{(1)}_{\phi \hat \phi} + \dots   
\end{equation}
 We then see $K\upp 1$ must scale as ${{\hat b\ll K \sqd \cc \L\sqd}\over{ |\hat \phi|\uu 3}}$.
Higher terms in the K\"ahler potential can be determined iteratively from the fixed point equation; this generates a series in the dimensionless ratio
$\hat b\ll K \L\sqd / |\hat \phi|\uu 3$.  All the basic ideas in the determination of the $\L$-dependent terms are as in the toy model and the \(O(2)\) model.

\heading{Reduction to Goldstones in the supersymmetric model}

Now we write the leading terms in the large-$J$ effective action in terms of $ |\phi|$
and $\chi \equiv {\tt arg}(\phi)$ as in the \(O(2)\) model.
\begin{multline}
{\cal L} = \hat{b}\ll K {{(\pp \phi)(\pp\phb)}\over{|\phi|\uu \hh}} + V(|\phi|) = \hat{b}\ll K \cc |\phi|\uu{3\over 2}  \cc (\pp\chi)\sqd +\hat{b}\ll K \cc {{(\pp |\phi|)\sqd}\over{|\phi|\uu\hh}} + V(|\phi|) \\ + (\text{higher~derivatives})+ (\text{fermions})\ ,
\end{multline}
with \(V(|\phi|) \propto |\phi|\uu{9\over 2}\) .

For configurations with $|\phi|$ constant, the action is minimized when
\begin{equation}
 (\pp\chi)\sqd \propto  |\hat{\phi}|\uu 3  + \text{(higher~derivatives)} + \text{(fermions)}\,.
\end{equation}

Eliminating $|\phi|$ classically and rewriting the effective action in terms of $\chi$ alone, we have
a Goldstone effective Lagrangian with exactly the same structure as in the case of the \(O(2)\) model,
with the exception of the presence of the fermions:
\begin{equation}
  {\cal L} = b\ll\chi \cc |\pp\chi|\uu 3 +
  (\text{lower order in $|\pp\chi| $}) + \text{(fermions)} \,.
\end{equation}

\subsection{Decoupling of the fermions}\label{sec:phi3-decouple-fermions}

The fermions, however, decouple from the dynamics.  As we show in the Appendix, they
have a rest energy $E\ll 0$ of order $|\chd| = O(\sqrt{\r})$, and therefore are heavier than the cutoff
of our effective theory, and we integrate them out.  
The fermions obtain their large masses from the Yukawa couplings \rr{SUSYYukawa}, which 
 allow quanta of R-charge carried by the fermions to convert into
quanta of R-charge in the $\dot{\chi}$ sector.  This induces a chemical potential for the fermions.

Importantly, the same Yukawa couplings also impart effective Majorana mass terms to the fermions that generate the gap.
For most purposes of the large-$J$ expansion of the lowest-dimension operators, the key fact is that the low-lying large-$J$ sector is described by
exactly the same universality class -- a conformally-invariant effective field theory for the R-Goldstone
$\chi$ -- as describes the large-$J$ sector of the critical $O(2)$ model.

\heading{Rest energies and speeds of the fermions}

The heavy fermions do display a few interesting features at large $J$, however.  For one, their rest mass
obeys a precise identity at leading order in $J$,
\bbb
E\ll 0 = 2 \cc {{d\D\ll J}\over{dJ\ll R}} = {3\over 2} \cc {{d \D\ll J}\over{dJ}} =  {3\over 2}\cc |\chd|\ ,
\een{RestMassID}
which is necessary for Bose--Fermi degeneracy.  The supercharges $Q\ll \a$ map from a sector
of $R$-charge ${{2J}\over 3}$ to a sector of $R$-charge ${{2J}\over 3} - 1$.  They
do so by removing one quantum of $R$-charge from the Bose condensate and replacing
it with a fermion that is almost at rest.  Therefore Bose--Fermi degeneracy implies \rr{RestMassID}
to leading order in $J$.
This value is completely fixed by the spontaneously broken superconformal symmetry and R-symmetry; the determination of the Goldstone fermion mass in terms of the chemical potential is a generalization of the formula of \cite{Nicolis:2012vf,Watanabe:2013uya} to the case of a breaking pattern described by a super-coset instead of a bosonic coset.

The other interesting feature of the fermion dynamics has to do with their propagation speed.
Like the bosons, the fermions propagate slower than the speed of light.  One might have
imagined that the fermion and Goldstone speeds are forced by supersymmetry to be the same,
but this is not so: The fermions actually have speed $\pm \hh$ times the speed of light, rather
than ${1\over{\sqrt{2}}}$ as the Goldstones do:
\bbb
E\ll{\rm f}(p) = E\ll 0 \pm v\ll {\rm f}|p| + O({{|p|\sqd}\over {|\chd|}})\ ,
\een{FDR}
where
\bbb
 v\ll {\rm f} = \hh\ .
\een{FermiSpeed}

This dispersion relation, particularly the unfamiliar appearance of the negative velocity is striking and begs for further explanation.  Like the value of the speed of the Goldstones, the speed of the fermions is also dictated by
superconformal symmetry.  Consider that the state $\kket J$ is the lowest state with $R$-charge $3J/2$ and is thus is annihilated by both the energy-raising, R-charge-raising generators $Q\dag$ and the
energy-lowering, R-charge-raising superconformal generators $S\dag$.  However
the state is non-\textsc{bps} and therefore can be annihilated by neither $S$ nor $Q$.  Therefore
the states $Q \cc \kket J$ and $S\cc\kket J$ are nonvanishing, and have energies 
$\D\ll J \pm\hh$.

These states are single-fermion excitations on top of a bose condensate of $\phi$-charge 
$J - {3\over 2}$, so the rest mass $E\ll 0 = {3\over 2}\cc |\chd|$ compensates the loss
of ${3\over 2}$ of a unit of $\phi$-charge from the Bose condensate, modulo an error of order $J\uu{-\hh}$.  The $O(1)$ difference $\pm\hh$ can thus only come from the kinetic energy of the fermions,
which is $\pm\cc v\ll f \cc |p|$.  On the unit sphere, the role of $p$ in the dispersion relation \rr{FDR} is played by the eigenvalue of the Dirac operator, which for the $\ell = \hh$ mode has absolute value $1$.  It follows that $v\ll f$ can \emph{only} be equal to $\hh$ if superconformal symmetry is respected, and that both positive and negative velocities must be present!

The dynamics of the fermions, particularly the negative velocity modes, is quite interesting in
its own right.  Such modes are widely studied in condensed matter systems.  They 
are a generic feature of fermions with a chemical potential
and Majorana mass term, as in our system.  The dispersion relation \rr{FermDisp4D}, for
instance, precisely appears in famous work by Fu and Kane~\cite{Fu:2008zzb} on the
interface between a topological insulator and a superconducting material.  For a review
of such modes and their role in condensed matter physics, the reader is referred to~\cite{Hasan:2010xy}.

\section{Observables}\label{sec:observables}

\subsection{Classical approximation to large-$J$ states}\label{sec:class-approx}

We now compute the large-$J$ expansion of operator dimensions using radial quantization.
That is, we use the state-operator correspondence, and the fact that the dimension of
the lowest operator of charge $J$ is equal to the energy of the lowest state of charge $J$
on a \emph{unit sphere}.

At leading order, all expectation values of operators are given by evaluation of those
operators in the lowest-energy classical solution with charge $J$.
By standard manipulations in classical mechanics, one may see that the
lowest classical solution of a system with a global symmetry is always invariant under
a "helical" symmetry -- that is, a symmetry under a combined time translation and global
symmetry transformation.  In terms of the Goldstone field, this is simply the statement that the
lowest-energy classical solution with a given global charge always has the Goldstone $\chi$
varying exactly linearly in time at a rate that is spatially independent.

This simplifies our search for the classical solution representing the ground state. Starting from the Lagrangian density 
\(\mathcal{L} = b_\chi \left( \dot \chi^2 - \nabla \chi^2 \right)^{3/2} \), the conjugate momentum to the field $\chi$ is
\begin{equation}
  \Pi = \frac{\delta \mathcal{L}}{\delta \dot \chi} = 3 b\ll\chi \dot \chi \left( \dot \chi^2 - (\gg\chi)\sqd \right)^{1/2},
\end{equation}
in terms of which the Hamiltonian density reads
\begin{multline}
 \mathcal{H} = \frac{1}{3 \sqrt{6}} \left[ \frac{1}{b_\chi} \left(9 b_\chi^2 (\nabla \chi)^4 + 4 \Pi^2\right)^{3/2} - 27 b_\chi^2 (\nabla \chi)^6+36 \cc  \Pi^2 \cc (\nabla \chi)^2  \cc  \right]^{1/2}.
\end{multline}
In terms of $\chd$, the Hamiltonian density is given by
\begin{equation}
  {\cal H} = \left. b\ll\chi \left(\dot \chi^2 - (\gg\chi)\sqd \right)^{1/2} \left( \cc 2 \dot \chi\sqd + (\gg \chi)\sqd \cc \right) \right|_{\dot \chi^2 = \frac{(\nabla \chi)^2}{2} + \frac{\sqrt{4 \Pi^2 + 9 b_\chi^2 (\nabla \chi)^4}}{6 b_\chi}  } \ .
\end{equation}

For fixed $\dot{\chi}$ or fixed $\Pi$, the energy is monotonically increasing as a
funcion of $(\nabla\chi)\sqd$.  It follows that the spherically symmetric state is a global, rather than merely a local, minimum of the energy with fixed total charge.  Note that this is a rather different situation from that in the system studied in~\cite{Dyer:2015zha}, where the spherically symmetric classical solution for the lowest state at large monopole number is not even perturbatively stable.
In our case, the spherically symmetric state is not only a local minimum but a global minimum of the energy in the configuration space at fixed $J$.  It follows that the lowest state with sufficiently large charge always has spin zero in both the \(O(2)\) model and $W = \Phi\uu 3$ model.
%
%
For a spherically symmetric configuration, the Hamiltonian density of the lowest state of the system is
\begin{align}
\label{HamDens}
  \Pi &= \r =  3 b\ll\chi \chd\sqd = {J\over{4\pi }}\ , &
 {\cal H} &= {2\over{\sqrt{27b\ll\chi}}} \cc \Pi\uu{3\over 2}  = 
            \frac{J^{\frac{3}{2}}}{4 \pi \sqrt{27 \pi b_\chi}} 
\end{align}
and the classical solution is
\begin{align}
\chi\ll 0 &= \O\cc t, 
&
\Omega &\equiv \sqrt{J\over{12\pi b\ll\chi}}. \label{ClassSol}
\end{align}
The total energy of the state is the Hamiltonian density \rr{HamDens} integrated over $S^2$:
\begin{eqnarray}
E = \Delta(J) = {{J\uu{3\over 2} }\over{\sqrt{27\pi b\ll\chi}}}\ .
\end{eqnarray}

\subsection{Large-$J$ expansion of the operator dimension}\label{sec:large-J-op-dim}

\heading{Order $J\uu {+{3\over 2}}$ and $J\uu{+\hh}$ classical terms}

At order $J\uu{+{3\over 2}}$, the energy is just given by the classical energy of the lowest
classical solution. These contributions come directly from the only
two terms in the Lagrangian with positive $J$-scaling, the leading operator $\co\ll{{3\over 2}}
|\pp\chi|\uu 3$
and Ricci coupling $\co\ll{\hh} \cc {\tt Ric}\ll 3$.  As shown in Section~\ref{sec:OpClass}, there is no operator of order $J\uu 0$ that contributes classically to the energy of the bosonic ground state.  So the only order $J\uu 0$ terms come from the one-loop vacuum energy, calculated with the Gaussian action for fluctuations from the leading-order bosonic terms.  We now calculate the value of that correction to the vacuum energy.

\heading{Order $J\uu 0$ quantum correction}

We would like to use the leading large-$J$ terms in the conformal Goldstone action to calculate 
the energy of the lowest state.  Since the Goldstone action describes a large-$J$ universality class that is shared by the critical $O(2)$ model \rwa{and} the $W = \Phi^3$ model, the following discussion
shall apply to both cases together.

The action for the phase variable $\chi$ is given by
\begin{equation}
  \mathcal{L} = b\ll\chi \abs{\partial \chi}^3  = b\ll\chi \cc [- \pp\ll\m \chi\pp\uu\m\chi ]\uu{+{3\over 2}}\ .
\end{equation}
To quantize it, we break up the field as in Section \ref{sec:QuantumGoldstones}, \emph{i.e.},
\begin{equation}
\chi \equiv \chi\ll 0 + |\pp\chi\ll 0|\uu{-\hh} \cc \chh\ ,
\end{equation}
where $\chi\ll 0(t)$ is the classical solution \rr{ClassSol} and $\hat \chi$ is a canonically normalized fluctuation.

At large-$J$, we can Taylor expand the action:
\begin{equation}
  \begin{split}
    \mathcal{L} &= b\ll\chi \cc (\del_ t \chi \del_t \chi - \chi \bigtriangleup_{S^2}\chi)^{3/2}\\
    & = b\ll\chi \cc |\pp\chi\ll 0|\uu 3 + \frac{{3\cc b\ll\chi}}{2} \cc \chh \left(-\partial_t^2+{1\over 2}\bigtriangleup_{S^2}\right) \chh  + O\bigg ({{\chh\uu 3}\over{|\pp\chi\ll 0|\uu{3\over 2}}}\bigg ),
  \end{split}
  \label{NRGDispRel}
\end{equation}
where we have dropped a
total derivative term proportional to $\partial_\tau \chh$.

Wick rotating $t\to i \t$ and computing the determinant, we find
that the one-loop correction to the vacuum energy is given by the usual Coleman--Weinberg formula 
applied to a minimally coupled massless boson with propagation speed ${1\over{\sqrt{2}}}$ times the speed of light:
\begin{equation}
  \begin{split}
   {1\over{2T}}\cc \log\det\left(-\partial_\tau^2-{1\over 2} \cc \bigtriangleup_{S^2}\right)
    &= {1\over{2\sqrt{2}}} \cc \sum^\infty_{\ell=0}(2\ell+1)\sqrt{\ell(\ell+1)}.
  \end{split}
\end{equation}
(We have let $T\equiv \int\ll{-T/2}\uu{T/2} \cc d\t$ denote the total extent of (Euclidean) time.)
{The sum is divergent. To regularize it in terms of \(\zeta \) functions, we first rewrite it as
\begin{equation}
  I = \sum^\infty_{\ell=1}(2\ell+1)\sqrt{\ell(\ell+1)} = 2 \sum^\infty_{\ell=1}(\ell+\tfrac{1}{2})\sqrt{(\ell + \tfrac{1}{2})^2 - \tfrac{1}{4}} .
\end{equation}
This is the energy of a conformally coupled boson on a sphere with modes of energy \(\ell + \tfrac{1}{2}\) plus an extra mass term \(m^2 = -\tfrac{1}{4}\).
The mode expansion allows us to separate the convergent part from the divergent one (this is precisely a  heat-kernel regularization~\cite{Monin:2016bwf}):
\begin{equation}
  \begin{split}
    I ={}& 2 \sum^\infty_{\ell=1} \Big[ (\ell+\tfrac{1}{2})\sqrt{(\ell + \tfrac{1}{2})^2 - \tfrac{1}{4}} - ( (\ell + \tfrac{1}{2})^2 - \tfrac{1}{8}) \Big] \\
    & + 2 \sum_{\ell=1}^\infty \Big[ (\ell + \tfrac{1}{2})^2 - \tfrac{1}{8} (\ell + \tfrac{1}{2})^0 \Big] \\
    ={}& I_{\text{conv}} + I_{\text{div}} .
\end{split}
\end{equation}
The divergent part is a sum of Hurwitz zeta functions:
\begin{equation}
  I_{\text{div}} = 2 \zeta(-2, \tfrac{1}{2}) - \tfrac{1}{4} \zeta(0, \tfrac{1}{2}) - \tfrac{1}{4}  = -\tfrac{1}{4} ,
\end{equation}
while \(I_{\text{conv}}\) can be evaluated numerically.
}
Multiplying by ${1\over{2\sqrt{2}}}$, we have the one-loop renormalized vacuum energy of the bosonic $O(2)$ model in the
large-$J$ sector:
\begin{equation}
\D(J) \bigg |\ll{O(J\uu 0)~{\rm term}} = -0.0937256\ .
\end{equation}
Thus the energy of the lowest state has the asymptotic expansion
\bbb
\Delta(J) = c\ll{+{3\over 2}} \cc J\uu{+{3\over 2}}  + c\ll{+{1\over 2}} \cc J\uu{+{1\over 2}} 
- 0.0937256 + O(J\uu{-\hh})\ .
\een{AsypGroundEnergy}

It is sometimes helpful to re-express the structure of the asymptotic expansion in terms of
a sum rule.  We therefore characterize the expansion for the ground state energy \rr{AsypGroundEnergy} as the sum rule:
\begin{equation}
\label{SumRuleGroundEnergy}
  J\sqd  \cc \Delta(J) - \bigg ( {{J\sqd}\over 2}  + {J\over 4} + {3\over {16}} \bigg ) \cc \Delta(J-1)
- \bigg ( {{J\sqd}\over 2}  - {J\over 4} + {3\over {16}} \bigg ) \cc \Delta(J+1) =  0.035147 + O(J\uu{-\hh})\ .
\end{equation}
As we have emphasized throughout, there are no operators scaling as $J\uu 0$, so the order $J\uu 0$ term is universal.  

\heading{Noncontribution of higher-loop diagrams at order $J\uu 0$}

The Casimir contribution to the energy is the entire contribution
at this order. It is easy to see that there are no higher-order diagrams contributing to the energy
at order $J\uu 0$.  Once we write the action in terms of the canonically normalized fluctuation
field $\chh$, we see that every cubic vertex in a Feynman diagram scales as $|\pp\chi\ll 0|\uu{-{3\over 2}} \propto \r\uu{-{3\over 4}}$ at the largest, and every quartic vertex as 
$|\pp\chi\ll 0|\uu{-3} \propto \r\uu{-{3\over 2}}$ at the largest.
Therefore, a two-loop vacuum diagram contributes as $J\uu{-{3\over 2}}$ and
smaller.

Corrections to the ground state
energy from subleading explicit terms \rr{Op1Ov2} are larger but still too
small to contribute at order $J\uu 0$: The operator
$\co\ll{\hh} = {\tt Ric}\ll 3 |\pp\chi|$ gives $({\tt Ric})\ll 3 \cc {{(\pp\chh)\sqd}\over{|\pp\chi\ll 0|\sqd}} \propto \r\uu{-1}$ when expanded to quadratic order in fluctuations.  It affects the dispersion relation
and Casimir energy at order $J\uu{-1}$.

\subsection{Energies of excited states}\label{sec:ex-states}

It is now simple to work out the energy of the 
excited states and their conformal representation theory, \emph{i.e.}, which states are primaries
and which are descendants.
From \rr{NRGDispRel}, we see that at leading order the Lagrangian is just that of 
the free fields, so the equation of motion is just
\begin{equation}
\chi\ll{,tt}=+\frac{1}{2}\nabla^2\chi.
\end{equation}
Note that the dispersion relation is nonrelativistic; the propagation speed of the Goldstone
boson is ${1\over{\sqrt{2}}}$ times the speed of light.  We will see momentarily that this
value is the only one consistent with the spontaneously broken conformal symmetry.

The dispersion relation in terms of
frequencies and spins is as usual on a sphere:
\begin{equation}
\omega\ll\ell=\sqrt{\frac{1}{2}\ell(\ell+1)}.
\end{equation}
The energies of the excited states are therefore the energy of the ground state,
plus the sum of a set of frequencies $\o\ll\ell$.  

We start by examining the first-excited states.
Take $\ell=1$ for example. The first excited states have one 
excitation of spin 1 that increases the operator dimension
by 1, which corresponds to acting with $\partial$
on the primary operator to make a descendant.
The speed of the Goldstone, $1/\sqrt{2}$
plays an essential role here because otherwise
there would be no spin-$1$ oscillator which would 
increase the anomalous dimension by 1.  The conformal raising operator $P\ll\m$, in
other words, is just the Goldstone mode at $\ell = 1$.  Therefore the criterion for a state to
be a conformal primary is that it has no $\ell = 1$ modes excited, with only the
modes of $\ell = 2$ or greater occupied.

We can now compute the weights of primaries.  Acting with a single $\ell = 2$ oscillator gives
an energy of $\D(J) + \sqrt{3}$, which is the lowest primary state above the large-$J$ vacuum.
This state must be primary, because there is no state of charge $J$ with energy $\D(J) + \sqrt{3} - 1$.
We can construct scalar primary states as well.  The lowest-energy way to do this is by taking the 
singlet in the symmetric product of two $\ell = 2$ representations.  Therefore the lowest excited
scalar primary is obtained by acting with two $\ell = 2$ oscillators, and therefore has energy $\D(J) + 2 \sqrt{3}$.  The lowest vector primary is obtained by acting with one $\ell = 3$ oscillator and
one $\ell = 2$ oscillator, and has energy $\D(J) + \sqrt{3} + \sqrt{6}$.  

All these states have subheading large-$J$ corrections coming from loop effects and explicit
operator corrections in the Lagrangian, of course.  These contribute to $E - \Delta(J)$
at order $J\uu{-{3\over 2}}$ and $J\uu{-1}$, respectively.

\section{Large-J analysis of some other models}\label{sec:other-models}


\subsection{4D theories and their anomaly terms}

Consider a \ac{cft} in four dimensions with a $U(1)$ global symmetry, in which Weyl anomalies are present in the underlying \ac{cft}.  
And suppose further that the large $J$ universality class of this \ac{cft} is described again by the
a four-dimensional version of the conformal Goldstone system studied in the previous sections.\footnote{Though we do not know of any controlled example of such a \ac{cft}, we can certainly imagine they may exist.  For instance, an asymptotically free $SU(2)$ gauge theory with two pairs of complex scalar fields, each transforming as a doublet, may have a phase transition as the bare mass-squared parameter is varied from positive to negative and the $U(1)$ global symmetry acting on the doublets goes from unbroken to spontaneously broken.  Assuming such a phase structure exists, then
the boundary between the two presumably has a description in terms of a four-dimensional version of the conformal Goldstone theories we have discussed in previous sections.} In such a \ac{cft}, the anomalies must express themselves consistently at the level of the large-$J$ effective dynamics.  The result
is that the anomaly coefficients control certain terms in the large-$J$ expansion of the \ac{cft}.  
In four dimensions, 
the $a$-anomaly directly dictates the coefficients of certain terms in the large-$J$ effective action in flat space; certain
curvature-dependent terms of order $J\uu 0$ in the large-$J$ effective action are also dictated by the Weyl anomaly coefficients, of both $a$- and $c$-type.

The analysis of the Wess--Zumino term for anomalous Weyl symmetry has been carried out in~\cite{Luty:2012ww} (see their Eq.~(2.8)) and the result is that
\begin{equation}
  \begin{split}
    S_{\text{WZ}}[g_{\mu\nu}, \chi; a,c] = \int d^4x \sqrt{-g} \Big\{
    &-a \Big[ {\tt ln}(|\pp\chi|)\cc E_4 (g) \\
    &\hspace{2em} + 4 ( R^{\mu\nu}(g) -12 g^{\mu\nu} R(g) )
    \Omega^{-2} \delta_\mu \Omega \delta_\nu \Omega \\
    &\hspace{2em}  +  4\Omega^{-3}(\partial \Omega)^2 \Box \Omega - 2
    \Omega^{-4}(\partial \Omega)^4 \Big]\\
    &+ c \cc {\tt ln}(|\pp\chi|)\cc W^2(g)\Big\},
  \end{split}
\end{equation}
where $g$ is the background metric, $E\ll 4$ and $W\ll{\m\n}{}\uu\r{}\ll\s$ are respectively the
Euler density and Weyl tensor, 
and for us,
\begin{equation}
\O \equiv |\pp\chi|.
\end{equation}

The point here is that the logarithm of $|\pp\chi|$ plays exactly the same role as the effective dilaton $\t$ in the Wess--Zumino anomaly term. When there is a supersymmetric moduli space of vacua, the logarithm of the overall scaling modulus plays the same role.
It would be interesting to learn from this substitution whether any interesting insights can
be gained from unitarity constraints on Goldstone scattering, perhaps generalizing the
recent celebrated proof of the $a$-theorem in four dimensions~\cite{Komargodski:2011vj}.

\def\FOURDDEFOUT{ 

\todo[inline]{and this I would eliminate}

It is, however, not completely clear how to do this. There are various ways of ending up with the classical vacuum manifold being a circle, which is what we want to achieve. One is a $g\sqd |\phi|\uu 4$ potential that
is UV completed as $g M \cc |\phi|\sqd \sigma + M\sqd \sigma\sqd$.  Another way
is to use an $SU(2)$ gauge group with two scalar doublets $\phi_p\upp i$ with
a $|\phi|\uu 4$ potential that is again UV-completed as $g M \cc |\phi|\sqd \sigma + M\sqd \sigma\sqd$
with a neutral scalar $\sigma$.  It is however not clear that either of these flows to an IR
fixed point.  Certainly there is no Wilson--Fisher fixed point in 4D, if you define that
as the strict $\e\to 0$ limit of the WF fixed point in $4-\e$ dimensions.  However, on the
other hand, there is no reason there can't be a fixed point of such a scalar theory at
finite coupling that is unrelated to the WF fixed point.  At any rate, these theories, if they do
indeed have fixed points, should have a large-$J$ behavior where the leading term
in the large-$J$ Lagrangian is $|\pp\chi|\uu 4$, and $J\sim |\pp\chi|\uu 3$.  Then
the leading scaling of the energy would be $E\sim J\uu{{4\over 3}}$.  This is, again, 
in contrast to the case of theories with flat directions, where the relationship goes as 
$E\sim J$, and the local relationship between order parameters is $T\ll{00} \sim J\ll 0 \cc ({\tt Ric})\uu{+\hh}$.

} 

\subsection{Monopoles and S--duality in three dimensions}

\paragraph{Dualities at large $J$.}

Like many known theories of a compact boson in three dimensions, the conformal Goldstone system has a duality transformation to an Abelian gauge theory in three dimensions, where the Noether
current maps to the monopole current, \emph{i.e.}, the Hodge dual of the field strength tensor, ${\cal J}\uu\m \to \hh \cc \e\uu{\m\n\s} (dA)\ll{\n\s}$.  Then the total Noether charge of a state maps to the magnetic flux on the sphere.
In the language of operators, the Noether charge of an operator in the original variables is the monopole number of the operator.

We can use coordinate- and Weyl-transformation properties in order to see how the field strength tensor can be expressed in terms of $\chi$. We take here a convention where 
$\epsilon_{012}=1$ and thus has Weyl weight 0. Also as $F_{\mu\nu}$ has Weyl weight 0 and
we are in three dimensions, so that $\sqrt{|g|}$ has weight $-3$.  At leading order in the derivative expansion, we find that Weyl-invariance, diffeomorphism covariance, and charge quantization uniquely determine the relation
\begin{equation}
\label{eq:chi-to-F}
F_{\mu\nu} = \sqrt{2} \abs{\pp \chi} (* d \chi)\ll{\m\n} =
\frac{1}{\sqrt{2}} \abs{\partial\chi} \sqrt{\abs{g}} \epsilon_{\mu\nu\sigma}\partial^{\sigma}\chi.
\end{equation}
The inverse relation is
\begin{equation}
  \del_\mu \chi = \frac{1}{\sqrt{2}} \abs{F}^{-1/2} (*F)_\mu =   \frac{1}{\sqrt{2}} \abs{F}^{-1/2}  \sqrt{\abs{g}} \epsilon_{\mu \nu \rho} F^{\nu \rho},
\end{equation}
where the numerical factors have been chosen so that
\begin{equation}
  \abs{F}^2 = \abs{\del \chi}^4 .
  \label{SDualNorm}
\end{equation}

The duality means that 
the effective Lagrangian for the field strength is immediately derived
from the leading Goldstone action in Eq.\eqref{LeadingGoldstoneAction}:
\begin{equation}
\mathcal{L}= b_\chi|F|^{3/2} + \dots.
\end{equation}
This is consistent with the fact that the Weyl weight of the Lagrangian is 3.
An immediate consequence of the unusual form of the action is that the dimension of the lowest-lying monopole operator scales 
as monopole number to the ${3\over 2}$, for large monopole charge.\footnote{The same
scaling has apparently been derived for operators of large monopole number 
in three-dimensional fixed points flowing from weakly-coupled gauge theory or Chern--Simons
matter theory.  We thank Ethan Dyer for communicating this result to us~\cite{DyerStanfordApril}.}

Note that the relation between ${\cal J}\uu\m$ and $F\ll{\m\n}$ is
exact, but the relationship between ${\cal J}\uu\m$ and $\pp\ll\m\chi$ depends on higher terms in the Lagrangian.  However the modifications have a sub-leading effect at large magnetic flux number: corrections
are suppressed by inverse powers of constant field strength, with numerators proportional to derivatives of the field strength.

\section{Conclusions}\label{sec:conclusions}

We have performed a renormalization group analysis proving that certain simple bosonic and supersymmetric systems are described at large charge density by a simple conformal Lagrangian
for Goldstone fields. 
In the limit
where the charge density is large compared to the infrared energy scale, the system is weakly coupled both classically and quantum mechanically,
and can be quantized straightforwardly in perturbation theory.  In the limit of large charge $J$, the leading large-$J$ expressions for all quantities, such as energies and anomalous
dimensions, are controlled by the leading terms in the effective action.  We do not know at present how to calculate the coefficients of those terms analytically from first principles, and we expect
them to differ between conformal field theories.  For instance, the value of $b$ describing the large-charge sector
of the $W = \Phi\uu 3 $ supersymmetric model likely differs from the value of $b$ describing the large-charge sector
of the three-dimensional XY model.  In even dimensions certain coefficients in the large-$J$ effective action can be
computed in terms of some intrinsic data of the \ac{cft} such as anomaly coefficients.

We have also computed the dimensions of excited primary states, and found that they are given up
to order $J\uu 0$ by energies of free oscillators in the $\l\geq 2$ spherical harmonics on $S\uu 2$.
 
There are a number of interesting questions to investigate in the future.

Since some features of the three-dimensional model may appear counterintuitive,
it would be interesting to compare them with the analogous properties
of various known conformal models in two dimensions. The complete
solvability of these models would give more confidence in the
consistency of our framework.

We may hope our framework is powerful enough to provide insights
in the large--\(J\) behavior of other strongly coupled \acp{cft}
which are in general not tractable with known methods. Natural examples are 2D
$\sigma$-models with degenerating cycles, non-supersymmetric
Chern--Simons-matter theory at finite rank and level~\cite{Aharony:2012nh, Aharony:2015pla}, and the \((2,0)\) theory at finite $N$.

A brief list of further interesting directions is given below.

\paragraph{Constraints on large-$J$ Lagrangian parameters.}

In some cases, one can constrain terms in the large-$J$ effective theory in terms
of the microscopic theory.  This is principally the case for four-dimensional theories, where
certain terms are dictated by anomaly coefficients.  It would be valuable to understand whether
this can be done more generally by matching correlation functions.  It would be nice, \emph{e.g.} to match the $|x|\uu{-4}$ term of a current two-point function in three dimensions with a current correlation
function in the large-$J$ effective theory, in order to calculate the $b\ll\chi$-coefficient in terms of
\ac{cft} data.  

\paragraph{Theories with moduli spaces and BPS operators.}

For theories with moduli spaces and (equivalently) an infinite ring of \textsc{bps} primary scalar operators,
our methods certainly cannot add anything to the prediction of the leading-order energies of the \textsc{bps} states, since their dimensions are dictated by an exact formula to be equal to the $R$-charge,
$\D = |J\ll R|$.  However large-$J$ methods can certainly shed light on related questions not
controlled directly by superalgebraic considerations, such as the anomalous dimensions
of states lying just above the \textsc{bps}
bound\footnote{See~\cite{Kovacs:2013una} for a related discussion of
  the large-\(J\) sector.}. 

\paragraph{Comparison with other results on the critical $O(2)$ model.}
One would like to see how our asymptotic formulae for the operator dimension may fit with
other approaches to understanding the $O(2)$ model at the critical point.
For one, the model can be simulated on a lattice and our results tested numerically.  Also,
much recent progress has been made on the conformal bootstrap both for
the three-dimensional critical \(O(N)\) models and the \(O(2)\) model, specifically~\cite{Kos:2015mba}, as well as for the \(W = \Phi^3\) theory~\cite{Bobev:2015vsa,Bobev:2015jxa},
and it would be desirable to see how such analytic bootstrap methods may
confirm and/or complement the results obtained in the present paper.

\subsection*{Acknowledgments}

  The authors would like to thank Ethan Dyer, Richard Eager, Juan Maldacena, Mark Mezei, Mauricio Romo, Kallol Sen, and Aninda Sinha for valuable discussions and correspondence.  The work of S.H. was supported by the World Premier International Research Center Initiative (\textsc{wpi} Initiative), \textsc{mext}, Japan and a Grant-in-Aid for Scientific Research (26400242) from the Japan Society for Promotion of Science (\textsc{jsps}).  The work of S.R. is supported by the Swiss National Science Foundation (\textsc{snf}) under grant number PP00P2\_157571/1.  D.O. and S.R. would like to thank the Kavli \textsc{ipmu} for hospitality during part of this work.

\appendix

\section{Fermions and supersymmetry at large R-charge}

In this section of the Appendix, we work out some general features of the situation in which
we have a theory with four supercharges and an exact R-symmetry, and we examine the lowest 
state at large R-charge and the energies of the fermionic excitations above it.  The main focus
of this section is to derive at the Lagrangian level the large gap in the fermion sector in the sector
of large $R$-charge.

\subsection{Quantum mechanics with four supercharges}\label{WZQM}

We study a quantum mechanical theory with two complex supercharges $Q\ll\a, Q\dag\ll\a$, where
we take $\a$ to run over the indices $\a\in\{\uparrow, \downarrow\}$.  The \textsc{susy} algebra
is
\begin{align}
  \{Q\ll\a, Q\ll\b\} &= 0 & \{Q\dag\ll\a, Q\dag\ll\b\} &= 0 &
  \{ Q\ll\a, Q\dag\ll\b\} &= 2 H\cc\d\ll{\a\b} \ .
\end{align}
For a chiral
multiplet $(\phi, \psi, F)$, the \textsc{susy} transformations are
\def\rrrt{{\sqrt{2}}}
\def\dllsk{\llsk\llsk}
\def\Fbar{\overline{F}}
\def\Fb{\overline{F}}
\def\twoeqline#1#2#3#4{{#1} &= {#2} &\dllsk\qquad {#3} &= {#4} }
\def\TTL#1#2#3#4{\twoeqline{#1}{#2}{#3}{#4}}
\begin{equation}
\begin{alignedat}{3}
\TTL
	 {Q\ll\a \cdot \phi}	{- i \rrrt \cc\psi\ll\a}
	 {Q\ll\a \cdot \phi }	0\ , 
	 \\
\TTL
	{Q\dag \ll\a\cdot \phi}		0
	{Q\dag\ll\a\cdot \phi}		{-i \cc \rrrt \cc \psi\dag\ll\a}
	 \\
\TTL
	{Q\ll\a \cdot \psi\ll\b}		{\rrrt\cc\e\ll{\a\b} \cc F}
	{Q\ll\a\psi\cdot \dag\ll\b}		{\rrrt\cc \d\ll{\a\b}\cc\phbd\ , }
	\\
\TTL 
	{Q\ll\a\dag \cdot \psi\ll\b}		{\rrrt\cc \d\ll{\a\b}\cc\phd\ , }
	{Q\ll\a\dag \cdot \psi\dag\ll\b}		{\rrrt\cc\e\ll{\a\b} \cc \Fbar}
	\\
\TTL
	{Q\ll\a \cdot F}				0
	{Q\ll\a\cdot \Fbar}			{- i \cc \rrrt \cc \e\ll{\a\b}\cc\dot{\psi}\dag\ll\b}
	\\
\TTL
	{Q\ll\a\dag \cdot F}				{- i \cc \rrrt \cc \e\ll{\a\b}\cc\dot{\psi}\ll\b}
	{Q\ll\a\dag\cdot \Fbar} 0.
\end{alignedat}
\end{equation}

\def\Wbar{\overline{W}}
\def\bal#1\eal{\begin{align}#1\end{align}}

The simplest model with this super algebra and field content is the 0+1-dimensional
Wess--Zumino model, which is defined by a K\"ahler potential $K(\phi,\phb)$ and
holomorphic superpotential $W$.  Taking the K\"ahler potential to be the flat, unit-normalized one $K(\phi,\phb) \equiv \phi\phb$, we have
\begin{equation}
L\ll{K} \equiv -4\cc Q\ll\uparrow \cdot Q\ll\downarrow \cdot  Q\dag\ll\uparrow \cdot Q\dag\ll\downarrow \cdot K(\phi,\phb),
= |\dot{\phi}|\sqd + |F|\sqd + i \psi\dag\ll\a\dot{\psi}\ll\a + ( \text{ total~derivative} ),
\end{equation}
\bal
L\ll W &\equiv - {i\over 2} \cc Q\ll\uparrow \cdot Q\ll\downarrow \cdot W(\phi) = - W\pr (\phi) \cc F + i
\cc W\prpr(\phi) \cc  \psi\ll \uparrow \psi\ll\downarrow
\\
L\ll{\Wbar} \equiv L\ll W\dag &=  - {i\over 2} \cc Q\dag\ll\uparrow \cdot Q\dag\ll\downarrow \cdot \bar{W}(\phb) = - \bar{W}\pr(\phb) \cc \Fbar + i \cc \Wbar\prpr(\phb)\cc \psi\dag\ll\uparrow \psi\dag\ll\downarrow\ .
\eal
The total Lagrangian is $L\equiv L\ll{K} + L\ll{W} + L\ll{\Wbar} = L\ll{\rm bos} + L\ll{\rm ferm}$,
where
\newcommand{\env}[2]{\begin{#1}#2\end{#1}}
\bal
L\ll{{\rm bos}} &= |\phd|\sqd + |F|\sqd - F \cc W\pr(\phi) - \Fbar \cc \Wbar\pr(\phb)
\\
L\ll{{\rm ferm}} & = i \psi\dag\ll\a\psd\ll\a + i \cc W\prpr(\phi) \cc \psi\ll\uparrow \psi\ll\downarrow
+ i \cc \Wbar\prpr(\phb) \cc \psi\dag\ll\uparrow \psi\dag\ll\downarrow.
\eal
After integrating out the auxiliary fields, we have
\begin{align}
  F &= \Wbar\pr(\phb),& \Fbar &= W\pr(\phi),
\end{align}
\begin{equation}
L\ll{{\rm bos}} = |\phd|\sqd - |W\pr(\phi)|\sqd .
\end{equation}
Now specialize to the case
\bbb
W(\phi) = {{g\uu{q-1}}\over{(q+1) \cc \sqrt{q}}} \cc \phi\uu{q+1}\ ,
\een{HomW}
where we have chosen our normalization for $g$ to simplify the formul\ae .
For the superpotential \rr{HomW}
\begin{align}
  F &= {{g\uu{q-1}\cc \phb\uu q}\over{\sqrt{q}}} &                                                   \Fbar &= 	{{g\uu{q-1}\cc \phi\uu q}\over{\sqrt{q}}}
\end{align}
\begin{align}
	L\ll{{\rm bos}}  &=  |\phd|\sqd - {{g\uu{2q-2}}\over{q\sqd}} \cc |\phi|\uu{2q} 
	\\
	L\ll{{\rm ferm}} &=  i \psi\dag\psd + i \cc \sqrt{q} \cc g\uu{q-1} \cc \phi\uu{q-1} \cc \psi\ll{\uparrow} \psi\ll{\downarrow} +  i \cc  \sqrt{q} \cc g\uu{q-1}\cc \phb\uu{q-1} \cc \psi\dag\ll{\uparrow} \psi\dag\ll{\downarrow}.
\end{align}
Writing our complex scalar $\phi$ in polar coordinates as in \rr{PolVars}, we have
\bal
	L\ll{{\rm bos}}  &=  a\sqd\chd\sqd - {{g\uu{2q-2}}\over q } \cc a\uu{2q} 
	\\
	L\ll{{\rm ferm}} &=  i \psi\dag\psd 
	+  i \cc  \sqrt{q} \cc g\uu{q-1}\cc a\uu{q-1} \cc \bigg ( \cc 
	\exp{i \cc (q-1) \cc \chi} \cc \psi\ll{\uparrow} \psi\ll{\downarrow} + \exp{-i \cc (q-1) \cc \chi} \cc  \psi\dag\ll{\uparrow} \psi\dag\ll{\downarrow} \bigg )\ .
\eal
We now make an additional redefinition to eliminate the nonderivative coupling of the
R-Goldstone $\chi$ to the fermions:
\bbb
 \hat{\psi} \equiv \exp{- {{i \cc (q - 1)}\over 2} \cc \chi} \cc\psi \ , \llsk\llsk 
 \hat{\psi}\dag \equiv \exp{+ {{i \cc (q - 1)}\over 2} \cc \chi} \cc\psi\dag\ ,
 \xxn
\psi = \exp{- {{i \cc (q - 1)}\over 2} \cc \chi} \cc \hat{\psi} \ , \llsk\llsk \psi\dag = \exp{+ {{i \cc (q - 1)}\over 2} \cc \chi} \cc \hat{\psi}\dag\ ,
\een{HattedFermionDefs}
in terms of which the fermionic Lagrangian is
\def\dpsh{\dot{\psh}}
\begin{equation}
L\ll{{\rm ferm}} =  i \psh\dag\dpsh + {{q-1}\over 2} \cc\chd\cc \psh\dag\psh
	+  i \cc  \sqrt{q} \cc g\uu{q-1}\cc a\uu{q-1} \cc \bigg ( \cc 
	 \psh\ll{\uparrow} \psh\ll{\downarrow} +   \psh\dag\ll{\uparrow} \psh\dag\ll{\downarrow} \bigg )\ .
\end{equation}
For fixed $\chd$, the frequency of the $a$-oscillator scales as $\chd$, so we would like to integrate it out it of
the system and obtain an effective theory of the other degrees of freedom.  Working classically, we integrate out $a$ by eliminating it classically from the Lagrangian by minimizing its energy at fixed $\chd$, giving
\bbb
(g\cc a)\uu{2q-2} = \chd\sqd\ ,  \llsk\llsk a = g\uu{-1} \cc |\chd|\uu{{1\over{q-1}}}
\een{aval}
so for states with $a$ in its ground state we have
\begin{align}
L\ll{\rm bos} &= {1\over{g\sqd}} \cc {{q-1}\over q} \cc |\chd|\uu{{2q}\over{q-1}}\ ,\\
L\ll{{\rm ferm}} &=  i \psh\dag\dpsh + {{q-1}\over 2} \cc\chd\cc \psh\dag\psh
	+  i \cc  \sqrt{q} \cc |\chd| \cc \bigg ( \cc 
	 \psh\ll{\uparrow} \psh\ll{\downarrow} +   \psh\dag\ll{\uparrow} \psh\dag\ll{\downarrow} \bigg )\ ,
\end{align}
which we now write as
\begin{equation}
L\ll{{\rm ferm}} =  i \psh\dag\dpsh + \m\cc \psh\dag\psh
	+  i \cc M \cc \bigg ( \cc 
	 \psh\ll{\uparrow} \psh\ll{\downarrow} +   \psh\dag\ll{\uparrow} \psh\dag\ll{\downarrow} \bigg )\ ,
\end{equation}
\begin{align}
\label{MAndMuVals}
  \m &\equiv {{q-1}\over 2} \cc\chd \ ,& M &\equiv \sqrt{q} \cc |\chd|\ ,
\end{align}
where $\m$ is a kind of pseudo-chemical potential for the $\psh$-charge and $\psh$ is a
symmetry breaking mass term that violates $\psh$-charge.  The parameter $\m$ can
only be understood as a true chemical potential in the limit $M\to 0$.  For fixed $q$ we are
not within the confines of such a limit, so $\m$ does not admit a consistent interpretation as a chemical potential in this context.  

The energy of the fermionic excitations is
\bbb
E = \sqrt{\m\sqd + M\sqd} = {{q+1}\over 2} \cc |\chd|\ .
\een{QMFermDisp}
There are two degenerate modes with this energy, one with each possible eigenvalue $\pm 1$ of
the internal $SU(2)$ magnetic quantum number.  Notice that the excitation energy of the fermion is parametrically large as a function of $|\chd|$.  So if the upper limit on our energy is some fixed scale $\Lambda$ independent of $J$, then we are justified in treating the fermionic degrees of freedom as frozen in their ground states.

\subsection{Lift to field theory}\label{WZ4D}

Now we lift the model to $3+1$ dimensions; this is just the four-dimensional Wess--Zumino
model of a single chiral superfield a homogeneous superpotential of the form \rr{HomW}.  This
model has a Landau pole for $q = 2$ and is non-renormalizable for $q \geq 3$, but at present
we are only considering semiclassical aspects of the theory, and we can define the model
quantum mechanically with a cutoff $\Lambda$ if desired.

The form of the Lagrangian density is exactly the same as that of the Lagrangian in the previous Section~\ref{WZQM}, except that the fermion kinetic term acquires a 
piece with spatial derivatives
\bal
{\cal L}\ll{{\rm bos}} &= - \phi\ll{,\m}\phb\uu{,\m} - |W\pr(\phi)|\sqd
\\
{\cal L}\ll{\rm ferm} &= - i \psb\G\uu\m\pp\ll\m\psi + i \cc W\prpr(\phi) \cc \psi\ll\uparrow \psi\ll\downarrow
+ i \cc \Wbar\prpr(\phb) \cc \psi\dag\ll\uparrow \psi\dag\ll\downarrow\ .
\eal

Specifying to the superpotential \rr{HomW}, going to polar field variables as in \rr{PolVars}, and
defining hatted fermions as in \rr{HattedFermionDefs}, we have
\begin{equation}
L\ll{{\rm ferm}} = - i \bar{\psh}\G\uu\m \pp\ll\m\psh - {\cal A}\ll\m\cc \bar{\psh}\G\uu\m\psh
	+  i \cc M \cc \bigg ( \cc 
	 \psh\ll{\uparrow} \psh\ll{\downarrow} +   \psh\dag\ll{\uparrow} \psh\dag\ll{\downarrow} \bigg )\ ,
\end{equation}
\begin{align}
  {\cal A}\ll\m &\equiv {{q-1}\over 2} \cc \chi\ll{,\m}\ , & M &\equiv \sqrt{q} \cc g\uu{q-1}\cc a\uu{q-1}\ .
\end{align}

Now restrict to the lowest-energy state with given $R$-charge, for which the gradient of $\chi$ is
constant and purely in the time direction, and $a$ is constant as well, satisfying \rr{aval}.
Then letting $\m\in (0,A)$ and using $\psb \equiv \psi\dag\G\uu 0$ and $\G\uu A \G\uu 0 = - 
\G\uu 0 \G\uu A = + \s\uu A$ when acting on Weyl spinors of a definite chirality, we have
\begin{equation}
L\ll{{\rm ferm}} =  i \psh\dag\dpsh + i \psh\dag \s\uu A \pp\ll A \dpsh + \m\cc \psh\dag\psh
	+  i \cc M \cc \bigg ( \cc 
	 \psh\ll{\uparrow} \psh\ll{\downarrow} +   \psh\dag\ll{\uparrow} \psh\dag\ll{\downarrow} \bigg )\ ,
\end{equation}
with $M$ and $\mu$ given as before by \rr{MAndMuVals}.  The excitation energies of the
fermions are then
\bbb
E\ll\pm = \sqrt{(|p| \pm \m)\sqd + M\sqd}\ , \llsk\llsk |p|\equiv \sqrt{p\ll A p\ll A}\ .
\een{FermDisp4D}
The energy of a fermion at rest in the reference frame of the charge density, has
energy 
\bbb
E\ll 0 = {{q+1}\over 2} \cc |\chd|.
\een{FermRestEnergy}
For small momenta $|p| \muchlessthan |\chd|$, the dispersion relation is 
\begin{align}
E\ll\pm &\sim E\ll 0 \pm v\ll f |p|\ , & v\ll f &= \frac{q-1}{q+1}\,.
\end{align}

The same dispersion relation~\rr{FermDisp4D} holds when the theory is reduced to any lower dimension $D$, with spatial momenta taking values  in $D-1$ dimensional vectors.  On reduction to $0+1$ dimensions we recover~\rr{QMFermDisp}.

\subsection{Nontrivial K\"ahler potential}

It is clear that the parameters $M, \m$ in the dispersion relation, expressed in terms of $\chd$,
are independent of the coupling $g$.  It follows immediately that
they are independent of the normalization of the flat K\"ahler metric as well, because the
normalization of the flat K\"ahler metric can always be absorbed into a redefinition of $g$,
\emph{via} a rescaling of the fields of the multiplet $(\phi, \psi, F)$.

It is slightly less obvious, but still easy to see,
that the value \rr{FermRestEnergy} still holds exactly, even if the
K\"ahler potential has a general homogeneous form $K \propto |\phi|\uu{2\a}$.  By
redefining 
\bbb
\phi = \tilde{\phi}\uu{1\over\a}\ ,
\een{FieldRedef}
we obtain a quadratic K\"ahler potential $K \propto |\tilde{\phi}|\sqd$ and a superpotential $W\propto \tilde{\phi}\uu{{q+1}\over\a}$.
We then recover the same system as in sections \rr{WZQM},\rr{WZ4D}, with $\chi$
replaced by the phase $\tilde{\chi}$ of $\tilde{\phi}$, and with $q$
replaced with 
\bbb
q\ll{{\rm new}} = {{q+1}\over\alpha} - 1\ .
\een{newq}
All formulae for the fermion dynamics from section \ref{WZ4D} then hold, with the 
value of $a$ replaced by $q\ll{\rm new}$.
 In the case of interest to us in the present paper, namely the three-dimensional superconformal
 fixed point, the exponent in the K\"ahler potential is $\a = {3\over 4}$, and the superpotential is
 cubic, so
in our notation $q = 2$.  Thus using \rr{newq}, we have $q\ll{\rm new} = 3$.

We therefore have
\begin{align}
\m &= \dot{\tilde{\chi}}\ , & M &= \sqrt{3} \cc \dot{\tilde{\chi}}\ , \\
E\ll 0 &= 2 \cc |\dot{\tilde{\chi}}|\ , & v\ll f &= \hh\ .
\end{align}

\subsection{A general supersymmetric identity on the rest mass}

The formula for the energy $E\ll 0$ in \rr{FermRestEnergy} is directly
dictated by supersymmetry.  Even though all supercharges are realized nonlinearly, there
are no massless goldstini in the system.  This is because the supercharges
do not act on the fixed-$J$ Hilbert space; rather, they map from the Hilbert space with one
value of the $R$-charge to another.  As a result, the rest mass of the fermion is given not
by zero, but by the change in energy under the removal of a unit of $\phi$-charge.
Since the ratio between the normalization of $\phi$-charge to that of $R$-charge is 
${2\over{q+1}}$, the condition of Bose--Fermi degeneracy is
\begin{equation}
\label{RestMassId}
E\ll 0 =  {{d\Delta(J)}\over{dJ\ll R}} = {{q + 1}\over 2} {{d\Delta(J)}\over{dJ}}\ ,
\end{equation}
so that the energy of a zero-momentum fermion can exactly compensate the loss of one unit
of $R$-charge from the Bose condensate.

\printbibliography

\end{document}